%% file: jpsi_paper.VFinal.tex
\documentclass[12pt]{article}
\usepackage{epsfig}
\usepackage{epstopdf}
\usepackage{amsmath}
\usepackage{hhline}
\usepackage{amssymb}
\usepackage{times}
\usepackage{cite}
\usepackage{booktabs}
\usepackage{xspace}
\usepackage{units}
\usepackage{subfigure}
\usepackage{caption}
\usepackage{rotating}
\usepackage{multirow}
\usepackage[]{lineno}
\usepackage{color}

\newlength{\dinwidth}
\newlength{\dinmargin}
\begin{document}  
\newcommand{\pom}{{I\!\!P}}
\newcommand{\reg}{{I\!\!R}}
\newcommand{\slowpi}{\pi_{\mathit{slow}}}
\newcommand{\fiidiii}{F_2^{D(3)}}
\newcommand{\fiidiiiarg}{\fiidiii\,(\beta,\,Q^2,\,x)}
\newcommand{\n}{1.19\pm 0.06 (stat.) \pm0.07 (syst.)}
\newcommand{\nz}{1.30\pm 0.08 (stat.)^{+0.08}_{-0.14} (syst.)}
\newcommand{\fiidiiiful}{F_2^{D(4)}\,(\beta,\,Q^2,\,x,\,t)}
\newcommand{\fiipom}{\tilde F_2^D}
\newcommand{\ALPHA}{1.10\pm0.03 (stat.) \pm0.04 (syst.)}
\newcommand{\ALPHAZ}{1.15\pm0.04 (stat.)^{+0.04}_{-0.07} (syst.)}
\newcommand{\fiipomarg}{\fiipom\,(\beta,\,Q^2)}
\newcommand{\pomflux}{f_{\pom / p}}
\newcommand{\nxpom}{1.19\pm 0.06 (stat.) \pm0.07 (syst.)}
\newcommand {\gapprox}
   {\raisebox{-0.7ex}{$\stackrel {\textstyle>}{\sim}$}}
\newcommand {\lapprox}
   {\raisebox{-0.7ex}{$\stackrel {\textstyle<}{\sim}$}}
\def\gsim{\,\lower.25ex\hbox{$\scriptstyle\sim$}\kern-1.30ex%
\raise 0.55ex\hbox{$\scriptstyle >$}\,}
\def\lsim{\,\lower.25ex\hbox{$\scriptstyle\sim$}\kern-1.30ex%
\raise 0.55ex\hbox{$\scriptstyle <$}\,}
\newcommand{\pomfluxarg}{f_{\pom / p}\,(x_\pom)}
\newcommand{\dsf}{\mbox{$F_2^{D(3)}$}}
\newcommand{\dsfva}{\mbox{$F_2^{D(3)}(\beta,Q^2,x_{I\!\!P})$}}
\newcommand{\dsfvb}{\mbox{$F_2^{D(3)}(\beta,Q^2,x)$}}
\newcommand{\dsfpom}{$F_2^{I\!\!P}$}
\newcommand{\gap}{\stackrel{>}{\sim}}
\newcommand{\lap}{\stackrel{<}{\sim}}
\newcommand{\fem}{$F_2^{em}$}
\newcommand{\tsnmp}{$\tilde{\sigma}_{NC}(e^{\mp})$}
\newcommand{\tsnm}{$\tilde{\sigma}_{NC}(e^-)$}
\newcommand{\tsnp}{$\tilde{\sigma}_{NC}(e^+)$}
\newcommand{\st}{$\star$}
\newcommand{\sst}{$\star \star$}
\newcommand{\ssst}{$\star \star \star$}
\newcommand{\sssst}{$\star \star \star \star$}
\newcommand{\tw}{\theta_W}
\newcommand{\sw}{\sin{\theta_W}}
\newcommand{\cw}{\cos{\theta_W}}
\newcommand{\sww}{\sin^2{\theta_W}}
\newcommand{\cww}{\cos^2{\theta_W}}
\newcommand{\trm}{m_{\perp}}
\newcommand{\trp}{p_{\perp}}
\newcommand{\trmm}{m_{\perp}^2}
\newcommand{\trpp}{p_{\perp}^2}
\newcommand{\alp}{\alpha_s}

\newcommand{\alps}{\alpha_s}
\newcommand{\sqrts}{$\sqrt{s}$}
\newcommand{\LO}{$O(\alpha_s^0)$}
\newcommand{\Oa}{$O(\alpha_s)$}
\newcommand{\Oaa}{$O(\alpha_s^2)$}
\newcommand{\PT}{p_{\perp}}
\newcommand{\JPSI}{J/\psi}
\newcommand{\sh}{\hat{s}}
\newcommand{\uh}{\hat{u}}
\newcommand{\MP}{m_{J/\psi}}
\newcommand{\PO}{I\!\!P}
\newcommand{\xbj}{x}
\newcommand{\xpom}{x_{\PO}}
\newcommand{\ttbs}{\char'134}
\newcommand{\xpomlo}{3\times10^{-4}}  
\newcommand{\xpomup}{0.05}  
\newcommand{\dgr}{^\circ}
\newcommand{\pbarnt}{\,\mbox{{\rm pb$^{-1}$}}}
\newcommand{\gev}{\,\mbox{GeV}}
\newcommand{\mev}{\,\mbox{MeV}}
\newcommand{\WBoson}{\mbox{$W$}}
\newcommand{\fbarn}{\,\mbox{{\rm fb}}}
\newcommand{\fbarnt}{\,\mbox{{\rm fb$^{-1}$}}}
\newcommand{\nbarn}{\,\mbox{{\rm nb}}}
\newcommand{\dsdx}[1]{$d\sigma\!/\!d #1\,$}
\newcommand{\eV}{\mbox{e\hspace{-0.08em}V}}
\newcommand{\gevsq}{\,\mbox{GeV$^{2}$}}
\newcommand{\gevsqInv}{\,\mbox{GeV$^{-2}$}}
%
%
\newcommand{\qsq}{\ensuremath{Q^2} }
\newcommand{\et}{\ensuremath{E_t^*} }
\newcommand{\rap}{\ensuremath{\eta^*} }
\newcommand{\gp}{\ensuremath{\gamma^*}p }
\newcommand{\dsiget}{\ensuremath{{\rm d}\sigma_{ep}/{\rm d}E_t^*} }
\newcommand{\dsigrap}{\ensuremath{{\rm d}\sigma_{ep}/{\rm d}\eta^*} }

\newcommand{\dstar}{\ensuremath{D^*}}
\newcommand{\dstarp}{\ensuremath{D^{*+}}}
\newcommand{\dstarm}{\ensuremath{D^{*-}}}
\newcommand{\dstarpm}{\ensuremath{D^{*\pm}}}
\newcommand{\zDs}{\ensuremath{z(\dstar )}}
\newcommand{\Wgp}{\ensuremath{W_{\gamma p}}\xspace}
\newcommand{\WgpRec}{\ensuremath{W_{\gamma p, rec}}\xspace}
\newcommand{\tRec}{\ensuremath{|t_{rec}|}\xspace}
\newcommand{\NegtRec}{\ensuremath{-t_{rec}}\xspace}
\newcommand{\ptds}{\ensuremath{p_t(\dstar )}}
\newcommand{\etads}{\ensuremath{\eta(\dstar )}}
\newcommand{\ptj}{\ensuremath{p_t(\mbox{jet})}}
\newcommand{\ptjn}[1]{\ensuremath{p_t(\mbox{jet$_{#1}$})}}
\newcommand{\etaj}{\ensuremath{\eta(\mbox{jet})}}
\newcommand{\detadsj}{\ensuremath{\eta(\dstar )\, \mbox{-}\, \etaj}}

\newcommand{\mum}{\ensuremath{\mu m \xspace}}

\newcommand{\Mmumu}{\ensuremath{{m_{\mu\mu}}}\xspace}
\newcommand{\Mee}{\ensuremath{{m_{ee}}\xspace}}
\newcommand{\Mll}{\ensuremath{{m_{\ell\ell}}}\xspace}

\newcommand{\JPsi}{\ensuremath{{J/\psi}}\xspace}
\newcommand{\iPsi}{\ensuremath{{\psi}}\xspace}
\newcommand{\PsiPrime}{\ensuremath{{\psi(2S)}}\xspace}
\newcommand{\JPsiToee}{\ensuremath{{\JPSI \rightarrow e^+ e^-}}\xspace}
\newcommand{\JPsiTomumu}{\ensuremath{{\JPSI \rightarrow \mu^+ \mu^-}}\xspace}
\newcommand{\GammaGammaToee}{\ensuremath{{\gamma \gamma \rightarrow e^+ e^-}}\xspace}
\newcommand{\GammaGammaTomumu}{\ensuremath{{\gamma \gamma \rightarrow \mu^+ \mu^-}}\xspace}
\newcommand{\GammaGammaToll}{\ensuremath{{\gamma \gamma \rightarrow \ell^+ \ell^-}}\xspace}
\newcommand{\epToeXll}{\ensuremath{{e p \rightarrow e X \: \ell^+ \ell^-}}\xspace}

\newcommand{\XTrue}{ \ensuremath{\vec{x}_{{\rm true}} }\xspace}
\newcommand{\YRec}{ \ensuremath{\vec{y}_{{\rm rec}} }\xspace}
\newcommand{\A} { \ensuremath{ \bf A  }\xspace}
\newcommand{\Desc} [3]{{#1} energy data set {#2} {#3} cross section }
\newcommand{\glref}[1]{Gl.~(\ref{#1})}

\input commands.tex

\def\Journal#1#2#3#4{{#1} {\bf #2} (#3) #4}
\def\NCA{\em Nuovo Cimento}
\def\NIM{\em Nucl. Instrum. Methods}
\def\NIMA{{\em Nucl. Instrum. Methods} {\bf A}}
\def\NPB{{\em Nucl. Phys.}   {\bf B}}
\def\PLB{{\em Phys. Lett.}   {\bf B}}
\def\PRL{\em Phys. Rev. Lett.}
\def\PRD{{\em Phys. Rev.}    {\bf D}}
\def\ZPC{{\em Z. Phys.}      {\bf C}}
\def\EJC{{\em Eur. Phys. J.} {\bf C}}
\def\CPC{\em Comp. Phys. Commun.}

\begin{titlepage}

\noindent
\begin{flushleft}
{\tt DESY 13-058    \hfill    ISSN 0418-9833} \\
{\tt April 2013}                  \\
\end{flushleft}

\noindent

\noindent

\vspace{2cm}
\begin{center}
\begin{Large}

{\bf Elastic and Proton-Dissociative  Photoproduction of $\boldsymbol{\JPSI}$ Mesons at HERA }

\vspace{2cm}

H1 Collaboration

\end{Large}
\end{center}

\vspace{2cm}

\begin{abstract}
Cross sections for elastic and proton-dissociative photoproduction of $\JPSI$  mesons 
are measured with the H1 detector in positron-proton collisions at HERA.
The data   were collected at $ep$ centre-of-mass energies $\sqrt{s}\approx 318 \, \gev$ 
and $\sqrt{s}\approx 225  \, \gev$, corresponding to  integrated luminosities of $\mathcal{L} = 130 \pbarnt$ 
and $\mathcal{L} = 10.8  \pbarnt$, respectively.
The cross sections are measured as a function of the photon-proton centre-of-mass energy 
in the range $25< \Wgp <  110 \, \gev$. 
Differential cross sections $\mathrm{d}\sigma / \mathrm{d}t$, where $t$ is the squared four-momentum
transfer at the proton vertex, are measured in the range $|t| < 1.2 \, \gevsq$ for the elastic
process and $|t| < 8 \, \gevsq$ for proton dissociation. 
The results are compared to other measurements. 
The $\Wgp$ and $t$-dependences are parametrised using phenomenological fits. 

\end{abstract}

\vspace{1.5cm}

\begin{center}
Submitted to \EJC 
\end{center}

\end{titlepage}

%
%
%
\begin{flushleft}
  \input{h1auts}
\end{flushleft}
%

\newpage

\section{Introduction}
\begin{figure}[hhh]
\center
\includegraphics[height=4cm] {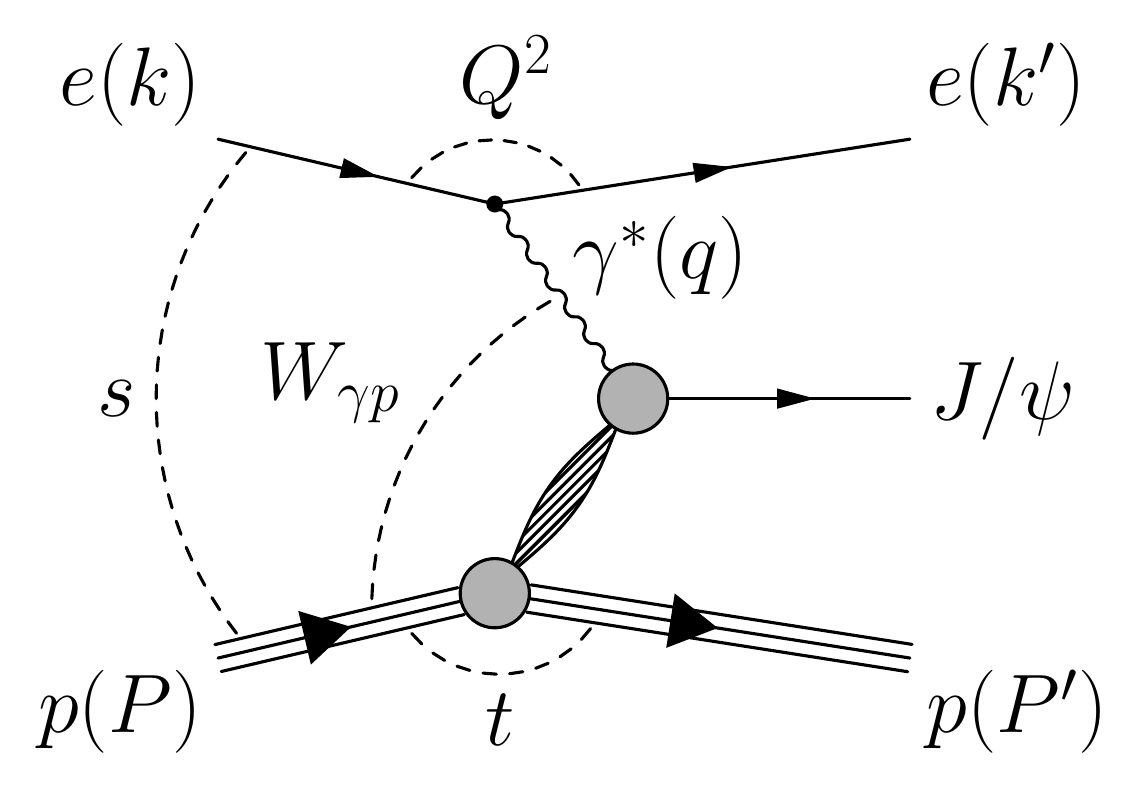}
\setlength{\unitlength}{4cm}
\put(-1.4,1.1) {\bf{a)}}
\hspace{2cm}
\includegraphics[height=4cm] {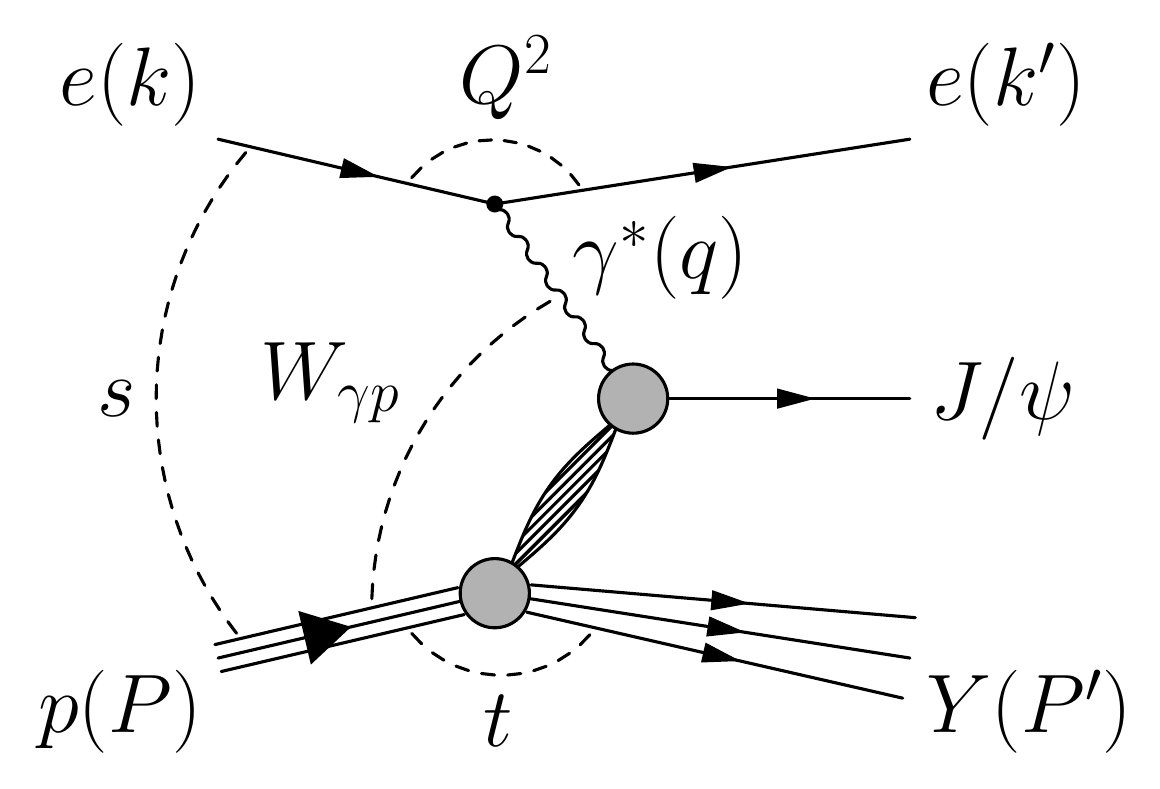}
\put(-1.4,1.1) {\bf{b)}}
\caption{Diffractive $\JPSI$ meson production  in  electron proton collisions:
 a) elastic $\JPSI$ production in which the proton stays intact and b) 
proton-dissociative $\JPSI$ production in which the proton dissociates to a low mass 
 excited  state with mass $M_Y>m_p$.}
\label{fig: FeynmanDiag} 
\end{figure}

This paper reports  a measurement of diffractive $\JPSI$ photoproduction 
in positron-proton interactions at HERA, $ep \rightarrow e \: \JPSI \: X$.
For the elastic regime $X$ denotes a proton, whereas for 
the proton-dissociative regime $X$ denotes
a proton-dissociative system $Y$ of mass $m_p < M_Y < 10 \gev $, as depicted
in figures~\ref{fig: FeynmanDiag}. \\

Diffractive vector meson production is characterised by the $t$-channel exchange of a colourless 
object between the incoming photon and proton. 
In the high-energy limit Regge theory
predicts~\cite{ReggeTheory, Donnachie:1994zb}   an approximate cross
section  dependence $\sigma \propto \Wgp ^ \delta$ as a function of the photon-proton 
centre-of-mass energy $\Wgp$.
For elastic production of light vector mesons ($\rho$, $\omega$, $\phi$) 
exponents  $\delta \approx 0.22$~\cite{Levy:1997pz} are observed.
In contrast, the cross section for elastic $\JPsi$ production, $\gamma p \rightarrow \JPSI \: p$, rises
more  steeply with 
$\Wgp$, $\delta \approx 0.7$~\cite{Aktas:2005xu, Chekanov:2002xi},  and is thus  incompatible with 
a universal pomeron hypothesis~\cite{Donnachie:1994zb}.
The $\Wgp$ dependence of  proton-dissociative $\JPSI$ 
production~\cite{Khoze:2006gg, Luna:2008pp, Gotsman:2007pn} is expected to be similar
to the elastic case.

Due to the presence of a hard scale,  the mass of the $\JPsi$ meson,  
calculations in perturbative Quantum-Chromo-Dynamics (QCD) 
are possible. The diffractive production of vector mesons 
can then be described in the proton rest frame by a process in which the photon fluctuates into a 
$q\bar{q}$ pair (or colour-dipole) at a long distance from the proton target. 
The $q\bar{q}$ pair interacts 
with the proton via a colour-singlet exchange, which in lowest order QCD is 
realised as a colourless 
gluon pair~\cite{Ryskin:1992ui,Brodsky:1994kf, Ivanov:2004vd, Brambilla:2010cs}. 
The steep rise of the cross 
section with $\Wgp$ is then related to the rise of the square of the gluon density towards 
low values of Bjorken 
$x$~\cite{Teubner:1999pm, Martin:2007sb, Diehl:2003ny, Frankfurt:2000ez, Caldwell:2010zza}.
 
The elastic and proton-dissociative $\JPSI$ cross sections as functions of 
the squared four-momentum transfer $t$ at
the proton vertex show a fast fall with increasing $|t|$~\cite{Aid:1996dn, Adloff:2000vm, Aktas:2005xu, Adloff:2002re, Breitweg:1997rg, Breitweg:1998nh, Chekanov:2002xi, Chekanov:2004mw,Aktas:2003zi, Chekanov:2009ab}. 
For the elastic $\JPSI$ cross section the $t$-dependence can 
be parametrised by an exponential function $\mathrm{d}\sigma / \mathrm{d} t \propto e^{- b_{el} |t|}$ 
 as expected from diffractive scattering. 
In an optical model the $t$-dependence of the elastic cross section carries information
on the transverse size of the interaction region. 
The proton-dissociative 
cross section falls less steeply than the elastic one and becomes dominant 
at $|t| \gtrsim 1\gevsq$.  
The  differential proton-dissociative cross section as a function of $t$ is parametrised
with a power-law function $\mathrm{d}\sigma /\mathrm{d}t \propto (1+(b_{pd}/n) |t|)^{-n}$,
which for low $|t|$  has an approximate exponential behaviour,  $\propto e^{- b_{pd} |t|}$. 

Diffractive $\JPSI$ production has been studied previously at HERA
at low values of $|t|$~\cite{Aid:1996dn, Adloff:2000vm, Aktas:2005xu, Adloff:2002re, Breitweg:1997rg, Breitweg:1998nh, Chekanov:2002xi, Chekanov:2004mw},
and also at very large values of $|t|$~\cite{Aktas:2003zi, Chekanov:2009ab},
where proton-dissociative $\JPSI$ production dominates. 

In this analysis cross sections are determined simultaneously
for the elastic and proton-dissociative regimes.
In addition to a measurement at the nominal $ep$ centre-of-mass
energy of $\sqrt{s} \approx 318 \gev$,  data recorded at a lower 
centre-of-mass energy of $\sqrt{s} \approx 225 \gev$
are  analysed.
This low-energy data set extends 
the kinematic region in $\Wgp$ into the transition region
between previous diffractive $\JPSI$ measurements at HERA and
fixed target experiments~\cite{Binkley:1981kv, Denby:1983az}.
The  elastic and proton-dissociative  cross sections
as functions of $t$ and $\Wgp$ are subjected to phenomenological fits,
together with previous H1 data~\cite{Aktas:2005xu, Aktas:2003zi},
and are compared with QCD based dipole models~\cite{Martin:2007sb}. 


\section{Experimental Method}
\label{sec:analysis}
\subsection{Kinematics}
\label{sec: Kinematics}
The kinematics of the processes $ep \rightarrow e \, \JPSI \, X$, where $X=p$ or $Y$
(depicted in figure~\ref{fig: FeynmanDiag}), are
described by the following variables: the square of the $ep$
centre-of-mass energy $s = (P+k)^2$, the square of the $\gamma p$
centre-of-mass energy $\Wgp^2 = (q+k)^2$, the absolute value of the four-momentum
transfer squared at the lepton vertex $\qsq = -q^2 = -(k-k')^2$
and of the four-momentum transfer squared at the proton vertex 
$t = (P-P')^2$. The four-momenta $k, k', P, P'$ and $q$ refer to 
the incident and scattered beam positron, the incoming and
outgoing proton (or dissociated system $Y$) and the exchanged
photon, respectively.  

In the limit of photoproduction, i.e $\qsq \rightarrow 0$, 
the beam positron is scattered at small angles and  escapes
detection. In this regime the  square of the $\gamma p$
centre-of-mass energy can be reconstructed via the
variable $\WgpRec ^2 = s \, y^{rec}$, where $y^{rec}$ is the 
reconstructed inelasticity, measured as
$y^{rec} = \left( E_{\JPSI}-p_{z, \, \JPSI} \right) / \left( 2 \, E_e \right)$. Here, $E_{\JPSI}$ and $p_{z, \, \JPSI}$
denote the reconstructed energy and the momentum along the proton beam direction ($z$-axis) 
 of the $\JPSI$ meson
and $E_e$ is the positron beam energy. Furthermore, the variable $t$
can be estimated from the transverse momentum of the $\JPSI$ in
the laboratory frame via the observable  
$t_{rec} =  - p^2_{T, \JPSI}$.
The reconstructed variables $\WgpRec$ and $t_{rec}$ are only 
approximately equal to the  variables $\Wgp$ and $t$, 
due to their definition and due to the smearing effects of
the detector.
In particular, $- p^2_{T, \JPSI}$  is systematically larger than $t$  for 
events with a value of $\qsq$ close to the upper boundary  of $2.5 \gevsq$ 
used in the analysis. 
In such events  the $\JPSI$  recoils against the
scattered beam positron in addition to the proton. The measurement presented here corrects for
this recoil effect 
by the unfolding procedure described below. 

\subsection{Monte Carlo models}
\label{sec: MC}
Monte Carlo (MC) simulations are used to calculate acceptances and  efficiencies for triggering,
track reconstruction, event selection, lepton identification and background simulation.
The elastic and proton-dissociative $\JPSI$ signal events are generated 
using the program DIFFVM~\cite{DiffVM}, which is based
on Regge theory and the Vector Dominance Model~\cite{VDModel}.
For $\JPSI$ production with proton dissociation a mass dependence of
$\mathrm{d} \sigma / \mathrm{d}M_Y^2 \propto f(M_Y^2) M_Y^{-\beta}$
is implemented in DIFFVM.
Here  $f(M_Y^2) = 1$  for $M_Y^2 > 3.6 \gevsq$, 
whereas for lower values of $M_Y^2$ 
the production of excited nucleon states  is taken into account explicitly. 
The description of the forward energy flow and the simulated $\Wgp$ and $t$ dependences 
are improved by weighting the MC samples
in $\Wgp$, $t$ and $M_Y$  according to a functional behaviour  motivated by the 
triple pomeron model~\cite{TriplePomeron} for the proton-dissociative case.
The reweighting model contains seven parameters, which are  adjusted to the data~\cite{fhuber}. 
QED radiation effects, which
are particularly relevant for $\JPsiToee$ decays, are simulated with the program PHOTOS\cite{Photos}.
The non-resonant di-lepton background is estimated using the GRAPE generator~\cite{GRAPE}, which simulates
electroweak processes $\epToeXll$. Possible
interference effects between di-lepton production via electroweak processes and
$\JPSI$ decays are ignored.

For all MC samples detector effects are simulated in detail with the GEANT 
program~\cite{Brun:1987ma}. The
MC description of the detector response, including trigger efficiencies, is adjusted using comparisons with 
independent data. 
Beam-induced backgrounds are taken into account by overlaying
the simulated event samples  with randomly triggered events.
The simulated MC events are passed through the
same reconstruction and analysis software as is used for the data.

\subsection{Detector}
\label{sec: Detector}
The H1 detector is described in detail elsewhere~\cite{H1det, Appuhn:1996na}.
Only those components
essential for this analysis are described here.
The origin of the right-handed H1 coordinate system is the nominal 
$ep$~interaction point, with the direction of the proton
beam defining the positive $z$~axis (forward direction). 
Transverse momenta are measured in the \mbox{$x$-$y$}~plane. Polar 
($\vartheta$) and azimuthal ($\phi$) angles are measured with respect to this frame of
reference. 

In the central region (\mbox{$15^\circ\!<\!\vartheta\!<\!165^\circ$}) the
interaction point is surrounded by the central tracking detector (CTD).
The CTD comprises two large cylindrical jet chambers
(CJC1 and CJC2) and a silicon vertex detector~\cite{pitzl}.
The CJCs are separated by a further drift chamber
which improves the $z$~coordinate reconstruction.
The CTD detectors are arranged concentrically around the interaction 
region in a uniform solenoidal magnetic field of \mbox{$1.16\ {\rm T}$}. The 
trajectories of  charged particles are measured with a transverse 
momentum resolution of $\sigma(p_T)/p_T \approx 0.2\% \, p_T/{\rm GeV} 
\oplus 1.5\%$.
The CJCs also provide a measurement of
the specific ionisation energy loss $\mathrm{d}E/\mathrm{d}x$ of charged particles
with a relative resolution of $6.5\%$ for long tracks.

The liquid argon (LAr) sampling calorimeter~\cite{Andrieu:1993kh} surrounds the tracking
chambers and has a  polar angle  coverage of 
\mbox{$4^\circ\!<\!\vartheta\!<\!154^\circ$}.
It consists of an inner electromagnetic section with lead absorbers and an outer 
hadronic section with steel absorbers.
Energies of electromagnetic showers are measured with a precision of
$\sigma(E)/E=12\%/\sqrt{E/\gev}\oplus 1\%$ and those of hadronic showers with
$\sigma(E)/E=50\%/\sqrt{E/\gev}\oplus 2\%$, as determined in test beam 
experiments~\cite{Andrieu:1994yn,h1testbeam}.  In the backward
region (\mbox{$153^\circ\!<\!\vartheta\!<\!178^\circ$}), particle energies are measured
by a lead-scintillating fibre spaghetti calorimeter (SpaCal)~\cite{Appuhn:1996na}. 

The calorimeters are surrounded by the muon system. The central muon detector (CMD) is integrated 
in the iron return yoke for the magnetic field and consists of $64$ modules, which are grouped 
into the forward endcap, the forward and backward barrel 
and the backward endcap and cover the range $4^{\circ}\leq\vartheta\leq171^{\circ}$.

Two sub-detectors situated in the forward direction are used in this analysis. 
These are the PLUG calorimeter, which is situated at $z=4.9 \, \mathrm{m}$, 
and consists of four double layers of scintillator and lead absorber, 
and the $z=28 \, \mathrm{m}$ station of the forward tagging system (FTS), which comprises
scintillator counters situated around the beam-pipe.

H1 has a four-level trigger system. The first level trigger (L1)
is based on fast signals from selected sub-detector components, which are combined and refined at
the second level (L2). 
The third level (L3) is a software based trigger using combined
L1 and L2  trigger information. 
After reading out the full event information events are reconstructed and
subjected to an
additional selection at a software filter farm (L4).
The data used for this measurement were recorded using  the Fast Track Trigger 
(FTT)~\cite{FTT} which, based on hit information provided by the CJCs,
reconstructs tracks with subsequently refined granularity at the first two trigger levels,
first in the \mbox{$x$-$y$}~plane at L1 and then in three dimensions at L2. 

For the data set taken at $\sqrt{s} \approx 318 \gev$  the luminosity 
is determined from the rate of the elastic QED Compton
process $ep \rightarrow e\, \gamma p$, with the positron and the photon detected
in the SpaCal~calorimeter, and the rate of Deep-Inelastic Scattering (DIS) events measured in 
the SpaCal~calorimeter~\cite{QEDComptonAnalysis}.
For the data set taken at $\sqrt{s} \approx 225 \gev$  the luminosity 
determination is based on the measurement of the 
Bethe-Heitler process $ep \rightarrow e\, \gamma p$
 where the photon is detected in a calorimeter located at $z = - 104 \, \mathrm{m}$ downstream of
the interaction region in the electron beam direction.


\subsection{Event selection}
\label{sec: Event selection}
The  measurement  is based on two data sets, both recorded with
a positron beam energy of $E_{e} = 27.6 \gev$.
The first
data set was taken in the years 2006 and 2007, when HERA was operated with 
a proton beam energy of $920 \, \gev$, resulting in a centre-of-mass energy of
$\sqrt s \approx 318 \, \gev$.
It corresponds to an 
integrated luminosity of $\mathcal{L} = 130 \pbarnt$. 
The second data set was recorded in the last months before the shutdown of
HERA in 2007, when the proton beam had a reduced 
energy of $460 \, \gev$, resulting in $\sqrt s \approx 225 \, \gev$.
This data set corresponds to an integrated
luminosity of $\mathcal{L} = 10.8 \pbarnt$.
These two samples will be referred to as high-energy (HE) and low-energy (LE)
data sets in the following.

Photoproduction events are selected by requiring the absence of a high energy electromagnetic
cluster, consistent with a signal from a scattered beam positron in the calorimeters.
Events with positrons detected in the SpaCal or LAr~calorimeter with
with energy above $8 \gev$ are rejected.
This limits the photon virtuality to $\qsq \lesssim 2.5 \, \gevsq$, resulting in
a mean virtuality of $\langle \qsq \rangle = 0.1 \, \gevsq$.

The triggering of  events relies on the online reconstruction of 
 exactly two oppositely charged tracks with transverse momenta 
$p_T> 0.8 \gev$ by the FTT. This condition is verified offline using  reconstructed tracks
based on the full CTD information in the polar range
$20^\circ < \vartheta < 165^\circ$.

Electrons from $\JPSI$ decays are identified using an electron 
estimator $D$~\cite{BeatuyPhotoproductionPaper},
which is based on energy deposits and shower shape variables in the LAr calorimeter and the 
specific ionisation energy loss $\mathrm{d}E/\mathrm{d}x$ measured 
in the CJCs. The estimator is defined
such that $D = 1$ for genuine electrons and $D = 0$ for background from pions. 
The selection of $\JPsiToee$ events is performed by
requiring a well identified electron with $D > 0.8$ in the polar range
$20^\circ < \vartheta_{e} < 140^\circ$, and  by observing a specific ionisation loss 
of the second track compatible with the electron hypothesis~\cite{fhuber}.

In the selection of $\JPsiTomumu$  events one muon candidate is identified
either in the calorimeter or in the muon system in the polar angle 
range of $20^\circ < \vartheta_{\mu} < 162.5^\circ$~\cite{fhuber}.
In order to reject misidentified $\JPsiToee$ events in this sample, 
the measured $\mathrm{d}E/\mathrm{d}x$ values of both tracks must be
incompatible with the electron hypothesis~\cite{fhuber}.
The signature of a $\JPsiTomumu$ event  can also be mimicked by a muon from a  cosmic 
shower passing through the detector.
The corresponding background is rejected by an acollinearity cut and
a cut on the timing information from the CTD\cite{fhuber}.

In order to suppress remaining non-$ep$ background, the event vertex, which is
reconstructed from the charged tracks in the event, is required to be within
$35 \, \mathrm{cm}$ of the nominal interaction point.
 
The summed squared energies of the SpaCal and LAr calorimeter clusters 
not related to the $\JPSI$ decay and
above $400\mev$ 
 have to satisfy the condition 
 $\sum_i E_i^2 < 2.5 \gevsq$. 
This 
requirement
reduces the remaining background from
proton-dissociative $\JPSI$ production with $M_Y > 10 \gev$ to
less than $2\%$ and from
inelastic $\JPSI$ production to the per-mille level~\cite{fhuber}. 

The di-lepton invariant mass distributions as reconstructed 
from the tracks for  the muon and the electron selection are shown in figure~\ref{fig: MassPeaks}
for both the HE and LE samples.
In all distributions the $\JPSI$ peak
at $\Mll \approx 3.1 \gev$ is clearly visible. 
The prominent tail of the mass peak in the 
$\JPsiToee$ channel towards low values of $m_{ee}$ is due to QED radiation losses 
and bremsstrahlung from the electrons,
reducing their momenta.
There is also background from non-resonant QED processes
$\epToeXll$.
Non-resonant diffraction contributes as a background to the muon channel due to
pions misidentified as muons. 
In contrast, the electron channel has negligible pion
contamination near the $\JPSI$ mass peak due to the superior background
rejection  of the electron selection.

\subsection{Signal determination}
\label{sec: Signal determination}
\subsubsection{$\boldsymbol{\JPsiTomumu}$}
For the muon decay
channel the number of reconstructed $\JPSI$ mesons is obtained
from the invariant di-muon mass distributions $\mmumu$  in bins
of $t_{rec}$ and $\WgpRec$. This is done by
fitting the sum of a Student's t-function describing the signal and
an exponential distribution for the non-resonant background 
with an extended binned log-likelihood fit
using the RooFit package~\cite{Verkerke:2003ir}.

The fit model has the form
\begin{linenomath*}
\begin{align}
  \label{eq:FitModel}
  f(N_\text{sig}, N_\text{BG}, \mmumu; \mu, \sigma, n, c) = N_\text{sig} \, p_\text{sig}(\mmumu; \mu, \sigma, n) 
  + N_\text{BG} \, p_\text{BG}(\mmumu; c)
\end{align}
\end{linenomath*}
with free shape parameters $\mu$, $\sigma$, $n$ and $c$ describing the
probability density functions of the $\JPSI$ signal $p_\text{sig}$
and of the background $p_\text{BG}$. The number of
signal and background events are given by $N_\text{sig}$ and
$N_\text{BG}$, respectively. The probability
density functions  are defined as
\begin{linenomath*}
\begin{align}
  p_\text{sig}(\mmumu) &= n_\text{sig}\left( 1 + \frac{r^2}{n}\right)^{-0.5(n+1)},\quad
  r=(\mmumu-\mu)/\sigma , \quad n>0\quad\text{and}\label{eq:PdfStudentT} \\
   p_\text{BG}(\mmumu) &= n_\text{BG} \, e^{-c\cdot \mmumu}.
\end{align}
\end{linenomath*}
The factors $n_\text{sig}$ and $n_\text{BG}$ are chosen such that the probability densities are
normalised to one
for both $p=p_\text{sig}$ and $p=p_\text{BG}$, in the  fit range $2.3 \gev < \mmumu < 5 \gev$. 
The small $\psi(2S)$ contribution is also included 
in the fit, modelled by a Gaussian.

The results of the fits to  the di-muon samples are
shown together with the data in   figure~\ref{fig: MassPeaks}.
 The fit yields to $29931 \pm 217$ $\JPsiTomumu$ events for the HE data set 
 and $2266 \pm 56$ $\JPsiTomumu$ events  for the LE data sets.

\subsubsection{$\boldsymbol{\JPsiToee}$}
\label{subsubsec: ElectronChannel}
\begin{figure}[hhh]
\centering
\setlength{\unitlength}{7.5cm}
\includegraphics[width=7.5cm] {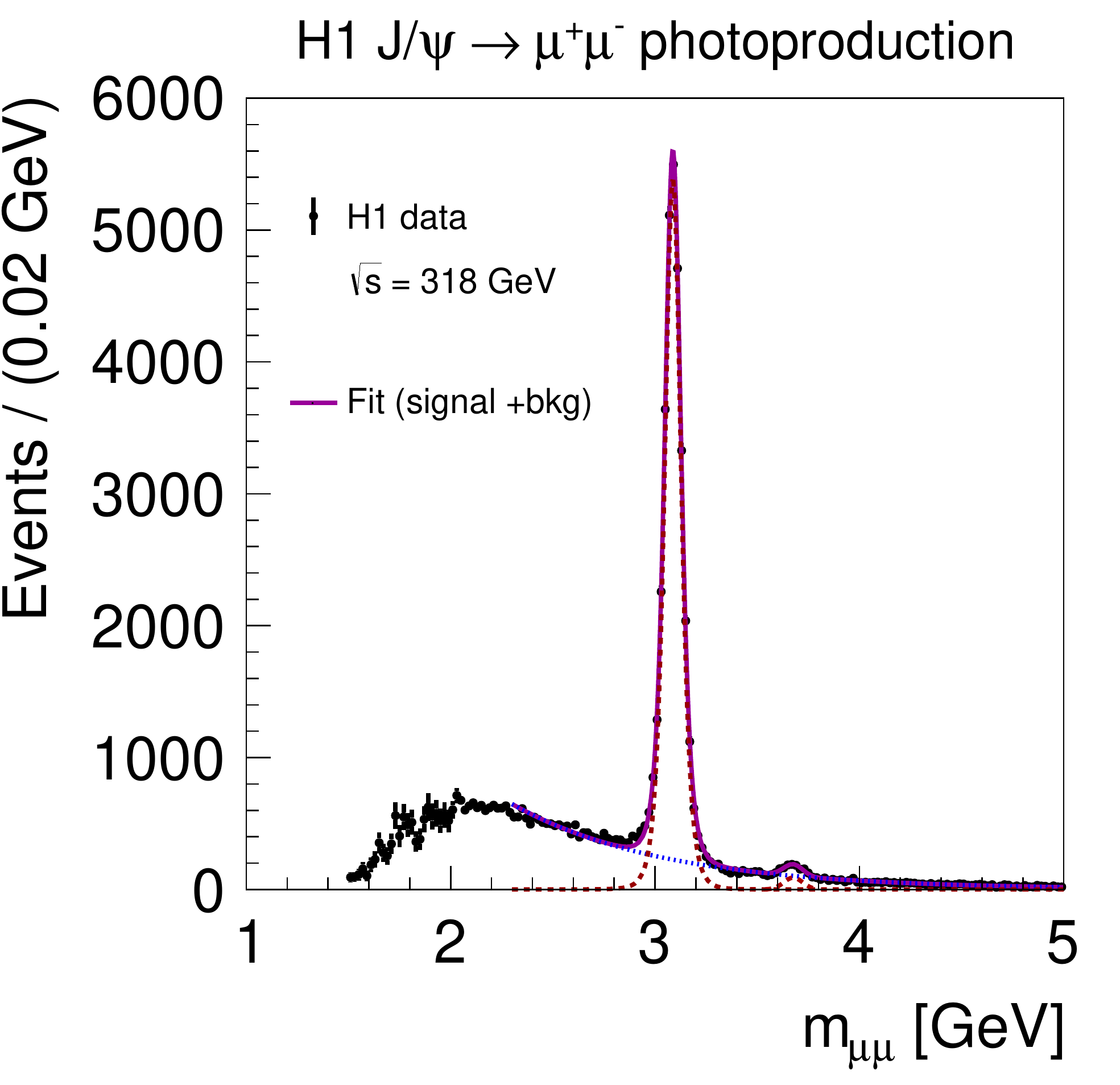} 
\put(-0.95, 0.95) {\bf{a)}}
\includegraphics[width=7.5cm] {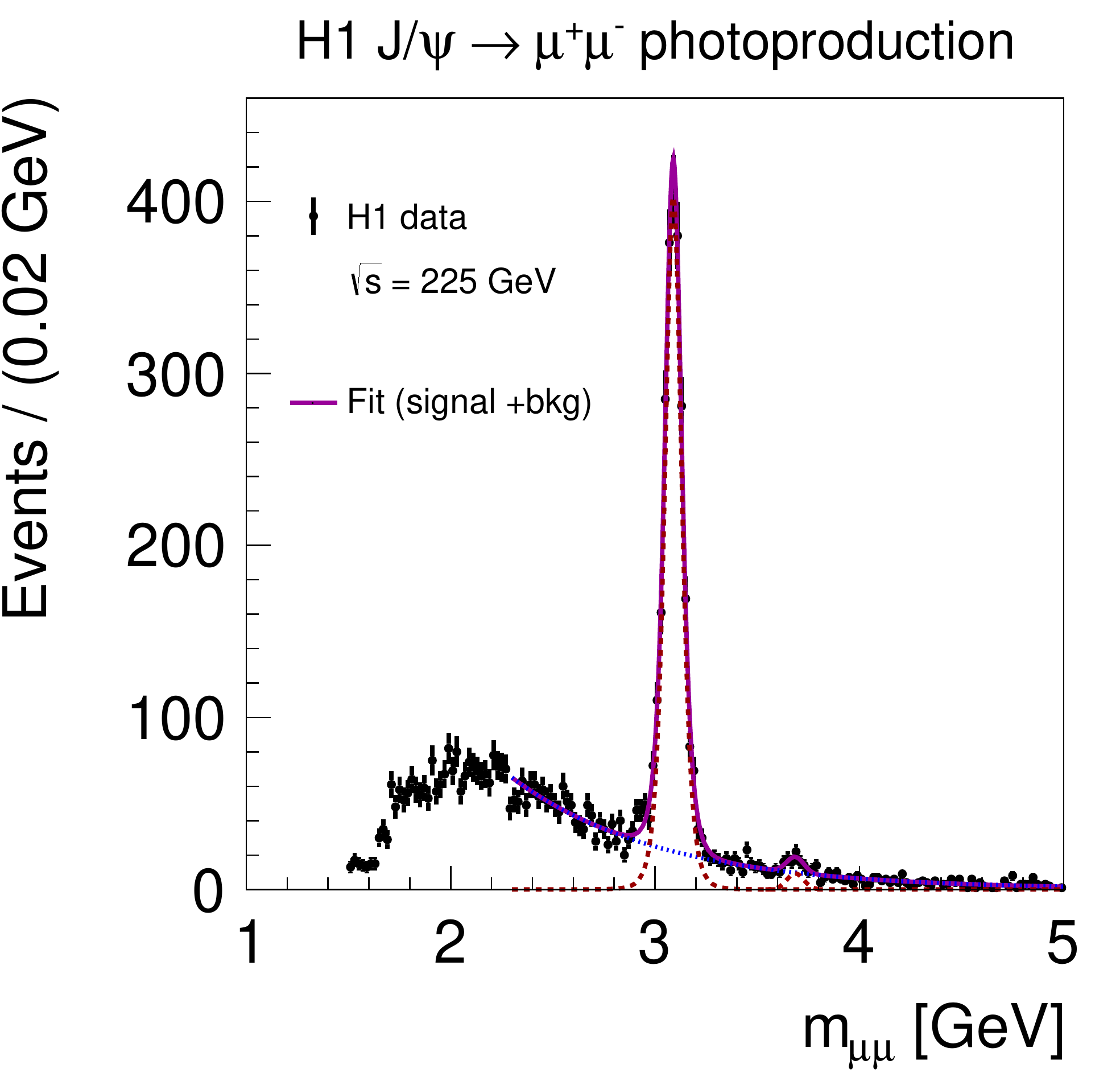} 
\put(-0.95, 0.95) {\bf{b)}} \\

\includegraphics[width=7.5cm] {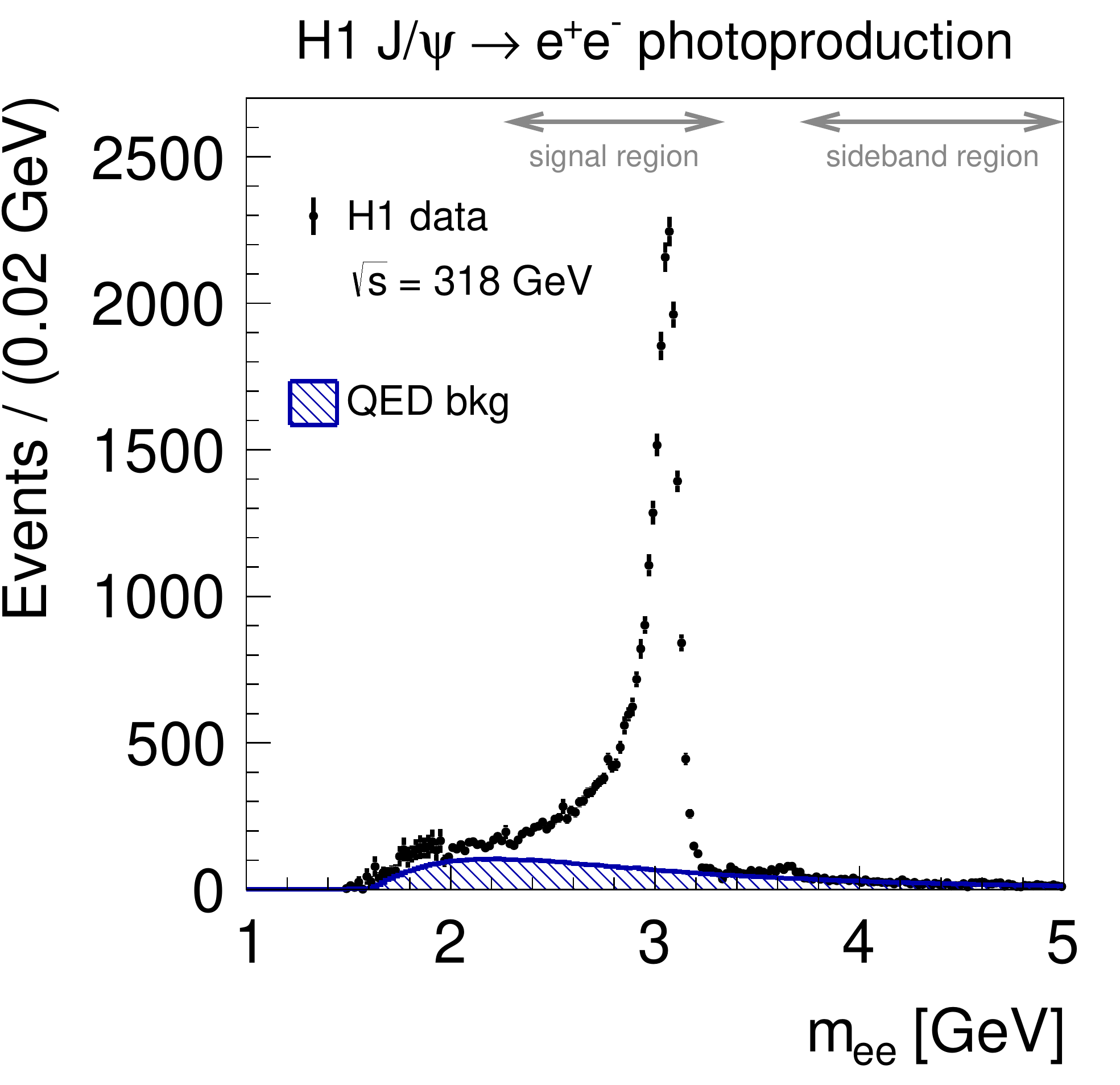} 
\put(-0.95, 0.95) {\bf{c)}}
\includegraphics[width=7.5cm] {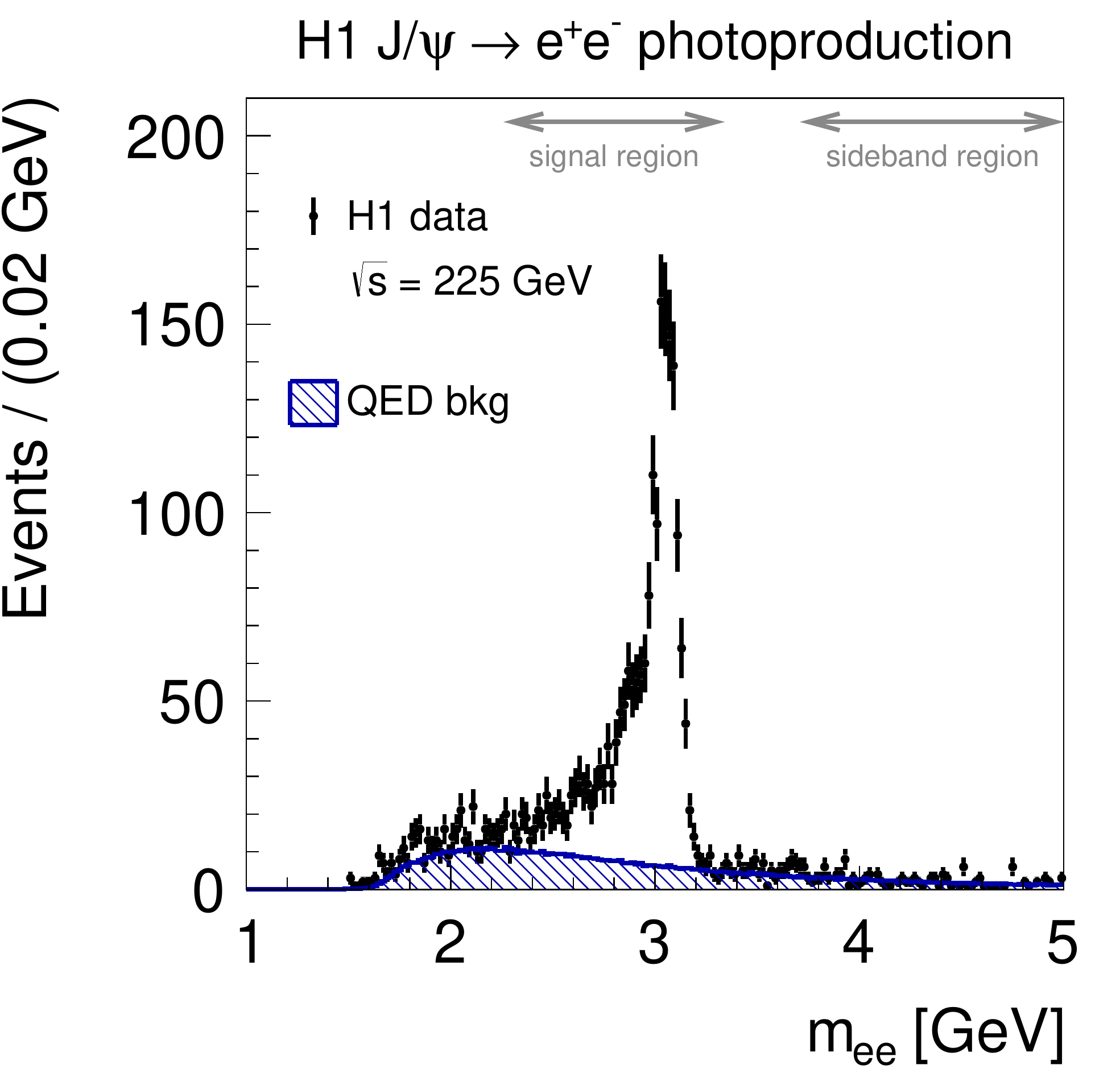} 
\put(-0.95, 0.95) {\bf{d)}}

\vspace{1cm}
\caption{Di-lepton invariant mass distributions for the high- and low-energy data sets in the
$\JPsiTomumu$ decay channel, figures~a) and~b), respectively, and 
for the $\JPsiToee$ decay channel, figures~c) and~d), respectively.
For the muon sample the  fits used to reconstruct the number of 
$\JPSI$ mesons are shown as well. For the electron sample the simulation
of the QED background $e p \rightarrow e X \: e^+ e^-$ is given by the shaded region and
the $\JPSI$ signal and sideband normalisation regions are  indicated.
}
\label{fig: MassPeaks} 
\end{figure}

For the electron decay channel the signal is
determined from the invariant di-electron mass distributions obtained in bins of $t_{rec}$
and $\WgpRec$. 
To reconstruct the number of $\JPsi$ mesons, a
different procedure from that used in  the muon channel is employed, which minimises the sensitivity
to details of  the large radiative tail of the  $\JPsi$ mass peak visible in
figure~\ref{fig: MassPeaks} and discussed above.
The non-resonant background, 
modelled using the  QED process $\epToeXll$ as simulated with GRAPE, 
is subtracted from the data.  This is possible due to the
negligible contamination from particles other than electrons at and 
above the $\JPSI$ mass peak.
The normalisation of the simulated QED background is determined 
prior to the background subtraction by fitting  the  background to the overall invariant
mass distribution in the mass window $ 3.75 < \Mee < 5\gev$ above the $\psi(2S)$ mass,
where only the QED  contribution is present. Within errors this normalisation factor
is consistent with unity. 

 After  background subtraction the remaining events are counted
within a window of $2.3 < \Mee < 3.3\gev$ around the nominal $\JPSI$ mass peak.
This yields  $23662  \pm 177$ $\JPsiToee$ events for the HE data set 
 and $1760 \pm 47$ $\JPsiToee$ events  for the LE data sets.
These  numbers of events are then corrected to account for the fraction of signal events
outside the counting window, which is close to $5\%$ as determined using the $\JPSI$ MC simulation. 
Within the counting window the $\JPSI$ MC simulation describes the behaviour
of the radiative tail in the data well.

\subsection{Experimental signatures of elastic and proton-dissociative processes}
\label{sec: tagging}
Proton-dissociative candidate events are identified by requiring either a large  
 value of $\tRec \gtrsim 1.5 \gevsq$ 
or 
  energy deposits in the H1 forward detectors, caused by  
fragments of the proton-dissociative system.
Three subdetectors, situated
at different locations, are used in this analysis to measure activity in the forward direction,
using the following requirements.
\begin{itemize}
 \item 	At least
  one cluster well above the noise level is found in the forward part of 
  the LAr, with 
  an energy above $400 \mev$ and $\vartheta < 10 ^ \circ$.
 \item	The summed energy of all clusters in the PLUG  calorimeter  
 	is above  $4\gev$, where all clusters above the threshold level of $1.2 \gev$ are considered.
 \item  Activity is observed in at least one  scintillator of the FTS station situated at $z = 28 \, \mathrm{m}$. 
\end{itemize}
If at least one of these conditions  is fulfilled,
the event is flagged as  tagged.  Identical tagging methods are
applied in the $e^+e^-$ and $\mu^+\mu^-$  channels.

\begin{figure}[hhh]
\centering
\setlength{\unitlength}{7.5cm}
\includegraphics[width=7.5cm] {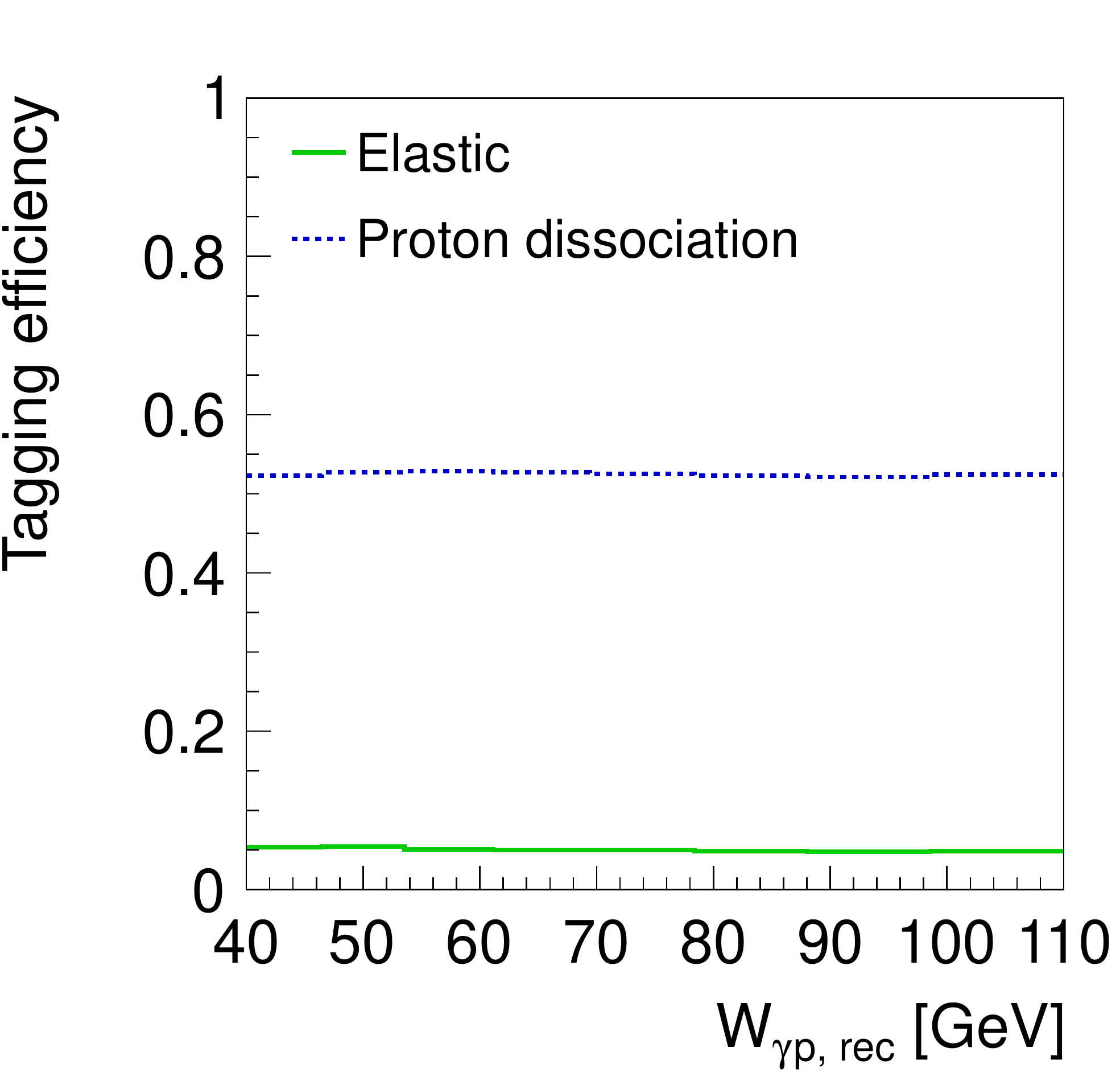}
\put(-0.95, 0.95) {\bf{a)}}
\includegraphics[width=7.5cm] {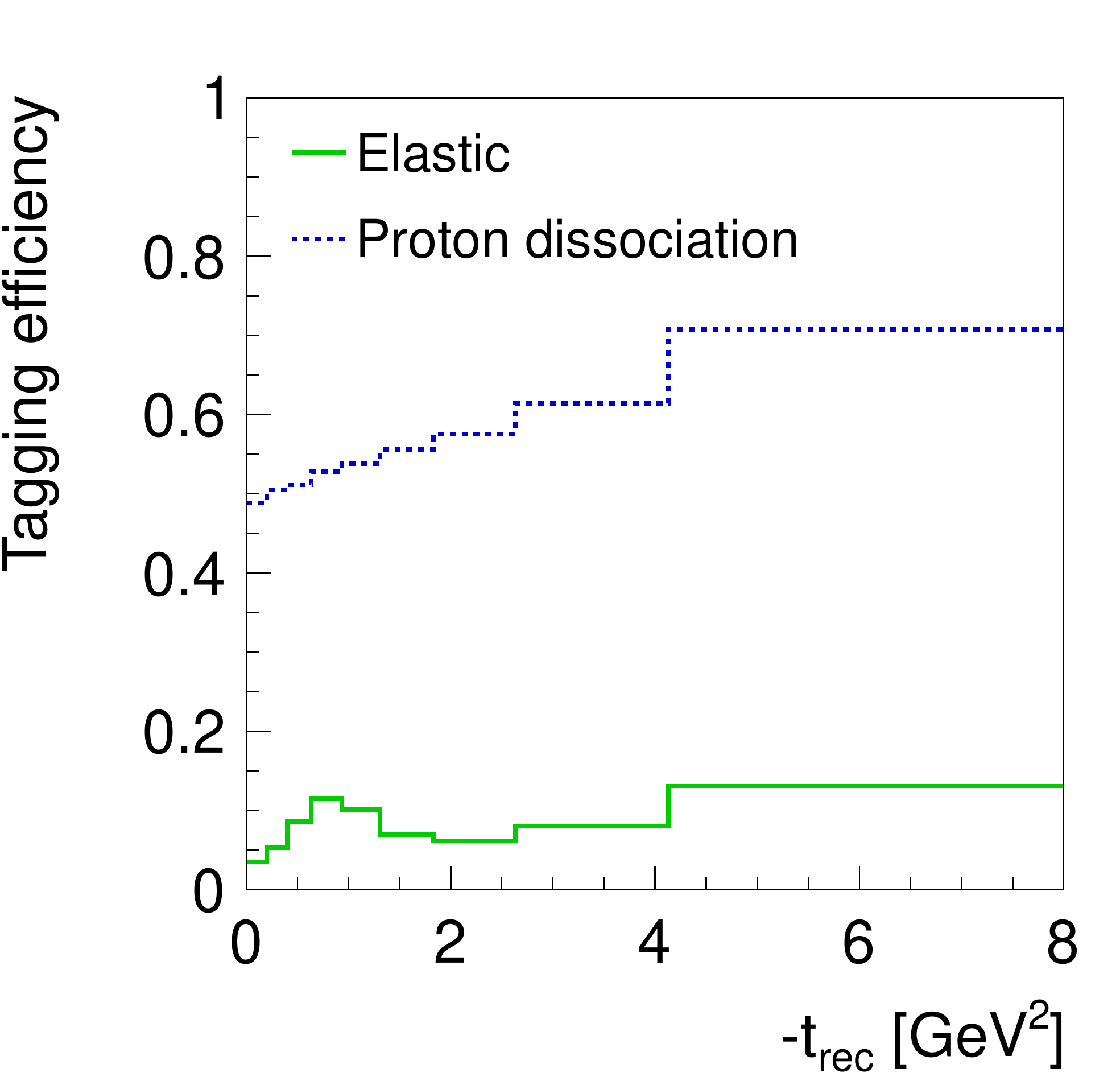}
\put(-0.95, 0.95) {\bf{b)}}\\
\includegraphics[width=7.5cm] {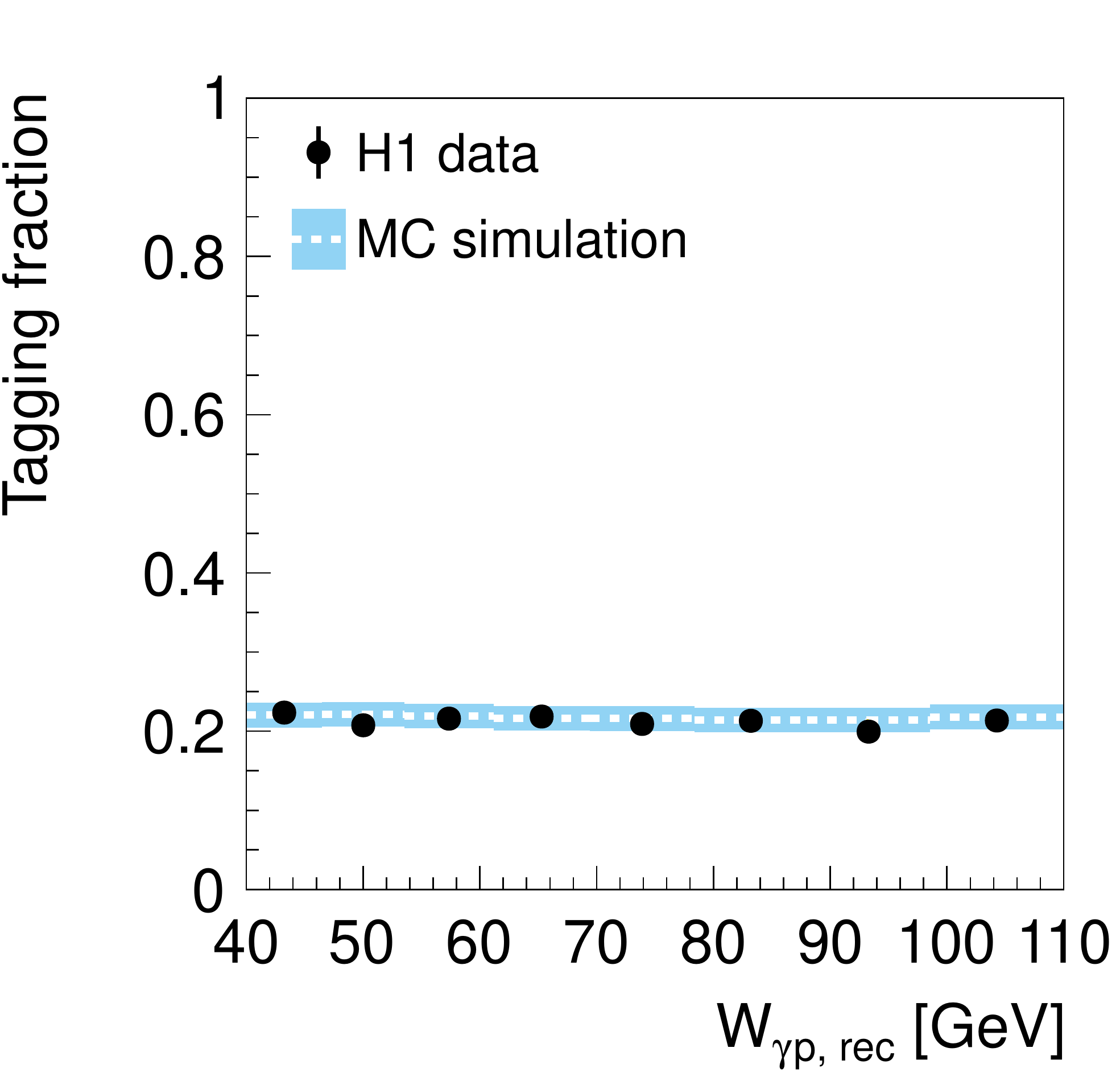}
\put(-0.95, 0.95) {\bf{c)}}
\includegraphics[width=7.5cm] {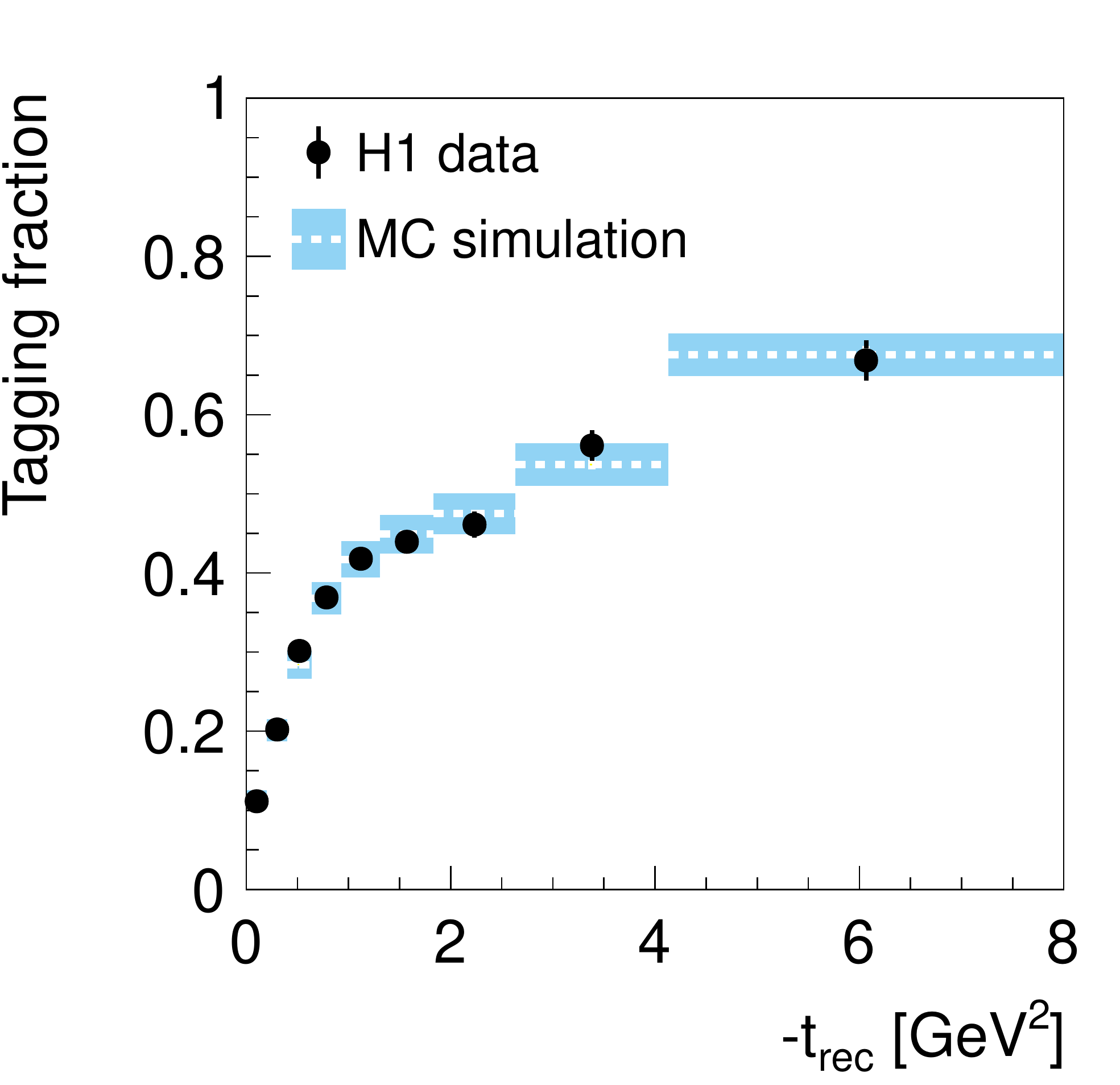}
\put(-0.95, 0.95) {\bf{d)}}

\vspace{1cm}

\caption{Tagging efficiencies as functions of (a) $\WgpRec$ and (b) $\NegtRec$
as obtained from the simulations of elastic and proton-dissociative 
$\JPSI$ production. 
Tagging fractions as functions of (c) $\WgpRec$ and (d) $\NegtRec$,
as obtained from the $e^+ e^-$ data set in the invariant mass window $m_{ee} = 2.3-3.3\gev$.
The data set contains elastic and proton-dissociative 
$\JPSI$ decays, as well as $e p \rightarrow e X \: e^+ e^-$ 
events. It is compared
to the simulation based on the event generators DIFFVM and GRAPE. 
The data (simulations) are shown by points (shaded bands).
The vertical spread of the bands represents the uncertainty
due to the tagging in the simulation.
}
\label{fig: Tagging} 
\end{figure}

In figure~\ref{fig: Tagging}  the simulated tagging efficiencies and tagging fractions 
observed in data and simulation are shown as functions of $\WgpRec$
and $\NegtRec$. The tagging
 fractions are obtained from the $e^+e^-$ sample, and contain QED contributions  in addition to 
di-electron events from diffractive $\JPsi$ production. 
In order to enrich it with genuine $\JPSI$
decays, the sample is restricted to invariant masses in the window $\Mee = 2.3-3.3\gev$. 
The tagging fractions observed in the data are compared to 
the simulation. The simulation is
based on the MC generators DIFFVM
for elastic and proton-dissociative $\JPSI$ production
 and GRAPE, which is used to describe the
QED background. The uncertainty in the simulation
due to the tagging of the forward energy flow is represented by the 
shaded bands. 
The tagging efficiency and  fraction show a flat behaviour as a function of $\WgpRec$.
A steep rise of the tagging fraction is observed as a function of $t_{rec}$, which reflects 
the relative elastic and proton-dissociative contribution in data.

An unambiguous event-by-event distinction between  elastic and proton-dissociative events 
is not possible with the H1 detector. Proton-dissociative
events can be misidentified as elastic events if the outgoing dissociated proton  
remains undetected due to the limited acceptance of the forward detectors. 
On the other hand, elastic events may have significant energy deposits in the forward
detectors due to possible beam induced background and may be misidentified
as proton-dissociative events. 
However, since the forward energy  flow is modelled by the MC simulation, 
elastic and proton-dissociative cross sections 
can be unfolded on a statistical basis.

\subsection{Unfolding}
\label{sec: Unfolding}

Regularised unfolding is used to determine  the elastic and 
proton-dissociative cross sections in bins of $t$ and $\Wgp$ from the number of events 
observed as a function of $t_{rec}$ and $\WgpRec$, respectively,  and from the
tagging information as described in the previous subsection.
The general procedure is described
in~\cite{BeatuyPhotoproductionPaper, PromptPhotons, DiffractiveDiJets} and the references therein. 
In the following only the aspects most relevant to this analysis
are summarised; further details are discussed in~\cite{fhuber}.

All efficiency corrections and migration effects are described
by a response matrix $\A$, which correlates the number of reconstructed
$\JPSI$ events in each analysis bin, represented by the vector $\YRec$, with 
the true distribution $\XTrue$ via the matrix equation $\YRec=\A \XTrue$.
The matrix element $A_{ij}$ gives the probability for an event originating from bin
$j$ of $\XTrue$ to be measured in bin $i$ of $\YRec$.
The unfolded "true" distribution  is obtained from the measured one by 
minimising a $\chi^2$-function $\chi^2(\XTrue; \YRec)$ by variation of $\XTrue$,
with a smoothness constraint determined by a 
regularisation parameter. This parameter  
is chosen such that the correlations in the covariance matrix of the unfolded
distribution $\XTrue$ are minimised. 

Two types of response matrix $\A$ are used: one to unfold differential
cross sections as a function of $t$, and one to unfold differential cross sections
as a function of $\Wgp$.
The response matrices are  calculated from the simulation and are defined such that the elastic and proton-dissociative
differential cross sections are determined simultaneously.
By using the tagging information for   small 
values of $\tRec \lesssim 1.5\gevsq$,
the elastic and proton-dissociative cross sections
are disentangled. 
Since the region of  large values of $\tRec$ 
is completely dominated by 
proton dissociation, no  tagging condition is applied.
Further, two reconstructed bins are associated with  each bin at the truth
level, in order to provide sufficiently detailed information on the probability distribution and to improve
the accuracy of the unfolding procedure.

The unfolding procedure is applied separately for the HE and the LE
data sets. The response  matrices 
for the LE data set are similar to those for
the HE case.  However they contain fewer bins
due to the smaller number of events.
 
\begin{figure}[hhh]
\centering
\setlength{\unitlength}{7.5cm}
\includegraphics[width=7.5cm] {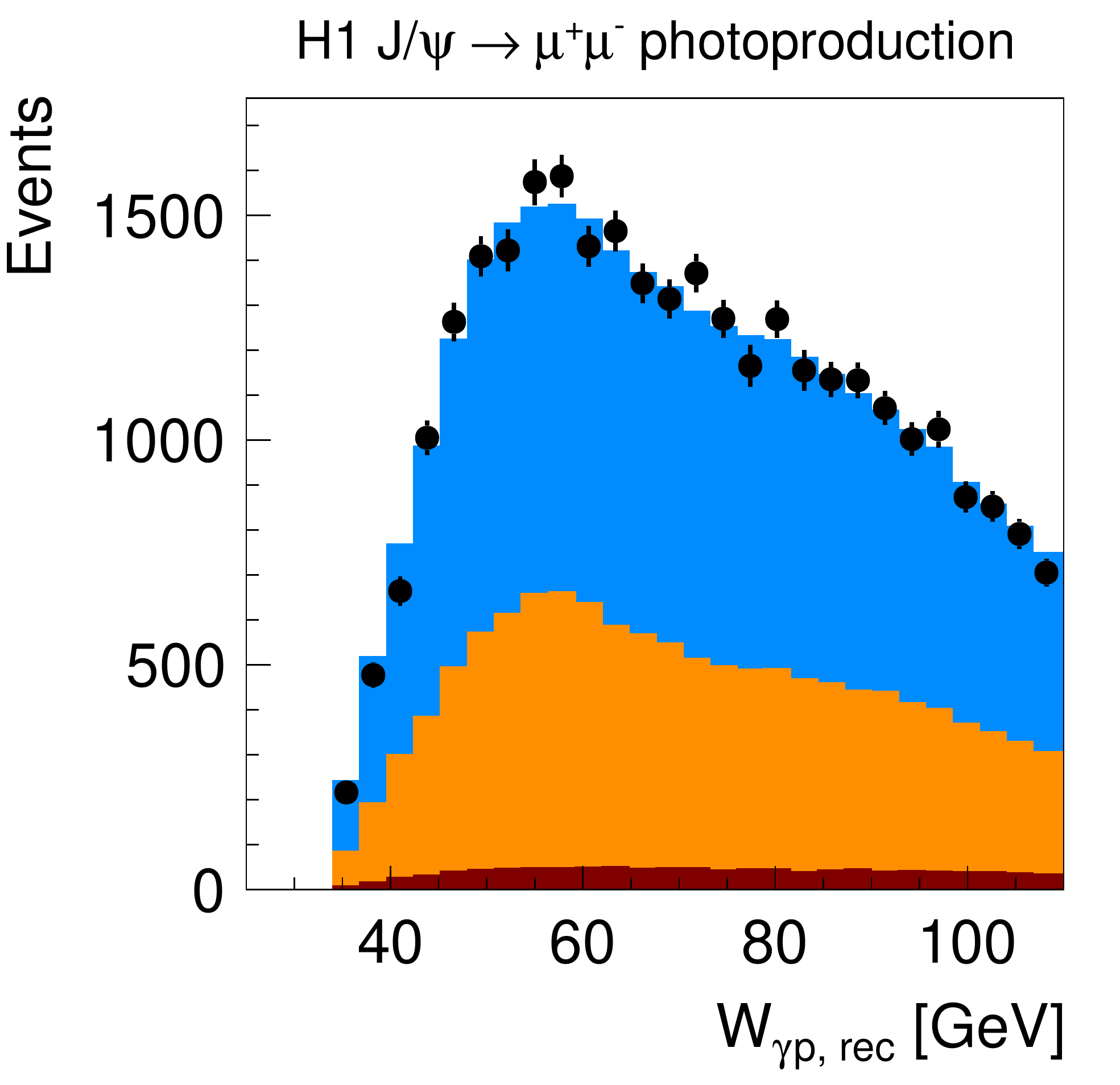}
\put(-0.95, 0.95) {\bf{a)}}
\includegraphics[width=7.5cm] {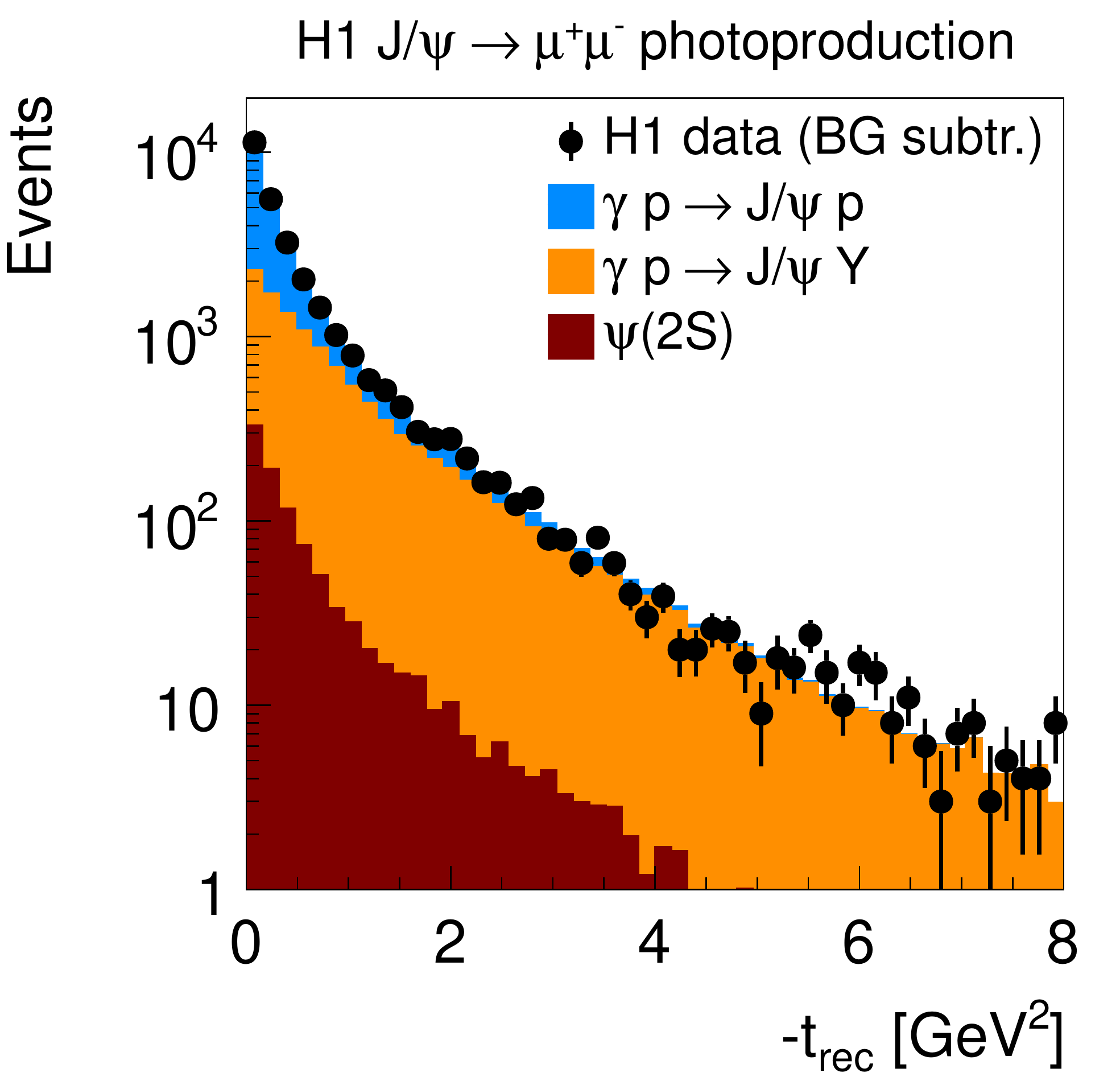}
\put(-0.95, 0.95) {\bf{b)}} \\

\includegraphics[width=7.5cm] {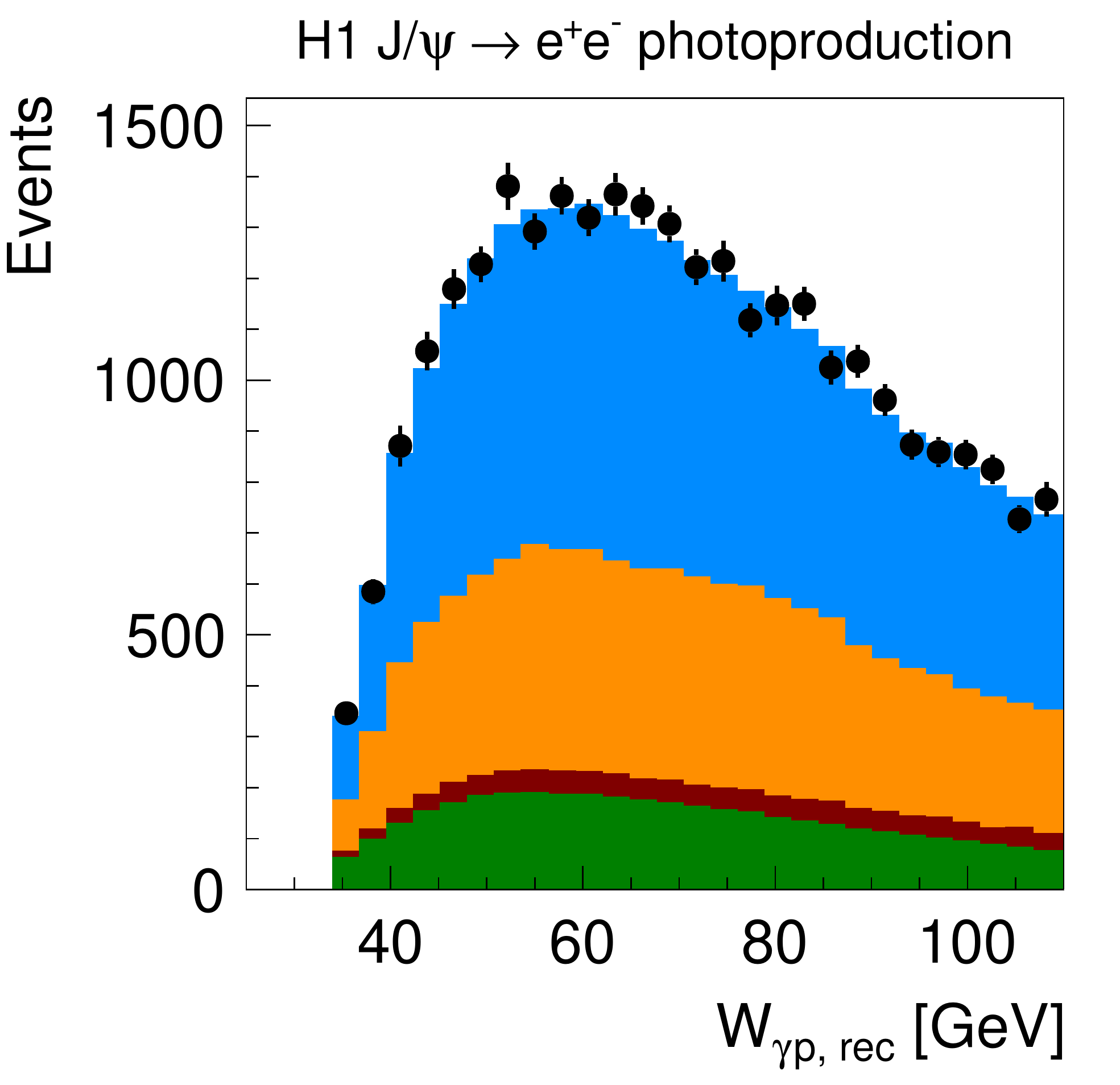}
\put(-0.95, 0.95) {\bf{c)}}
\includegraphics[width=7.5cm] {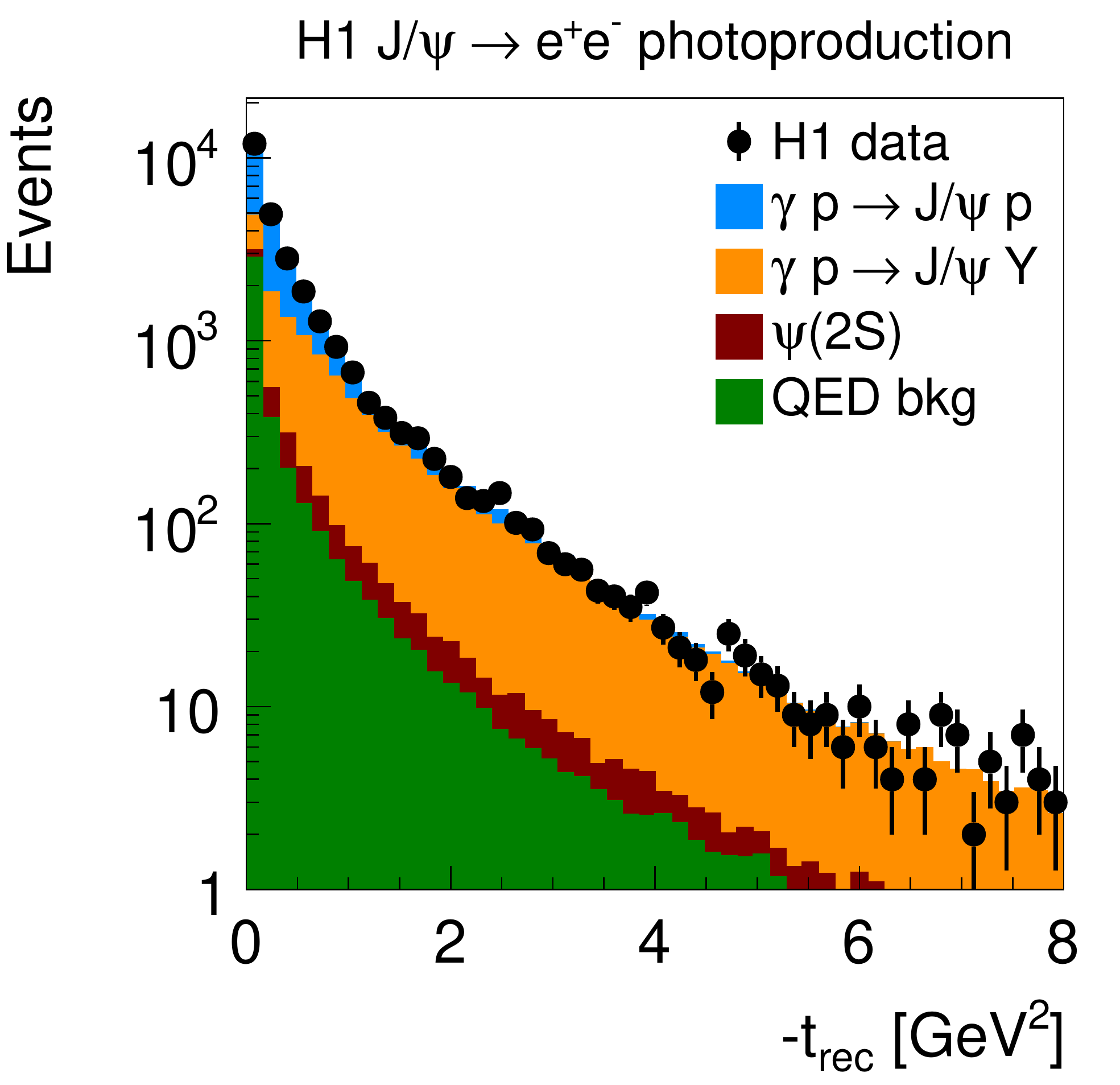} 
\put(-0.95, 0.95) {\bf{d)}} \\

\vspace{1cm}
\caption{
Observed distributions as functions of  $\WgpRec$ and $\NegtRec$
restricted in $\Mll$ to the $\JPSI$~signal region. 
The muon sample is shown
in~a) and~b), the electron sample is shown in~c) and~d).
The data, shown by the points, are compared to 
the simulation of elastic and proton-dissociative $\JPSI$ production.
Also shown is the contribution from $\psi(2S)$ events and, 
for the electron sample only, the QED background.
For the muon sample, background is subtracted from the data
using a sideband method.}
\label{fig: ControlDistributions} 
\end{figure}

In figure~\ref{fig: ControlDistributions} control distributions are shown for $\WgpRec$ and $\NegtRec$
separately for the $\mu^+\mu^-$ sample and the  $e^+e^-$ sample.
Both samples are restricted in $\Mll$ to the $\JPsi$ peak region, which is chosen
for the $\mu^+\mu^-$ sample to be $2.8 < \Mmumu < 3.3 \gev$. For the $e^+e^-$ sample
this region is enlarged to $2.3 < \Mee < 3.3 \gev$ in order not to cut into
the radiative tail.
The  relative fractions of the elastic and proton-dissociative 
events simulated with DIFFVM as determined in the unfolding procedure,
are also shown in
figure~\ref{fig: ControlDistributions}. The contribution from the
$\psi(2S)$ resonance is taken from the simulation, normalised using a previous
measurement~\cite{Adloff:2002re}. 
For the $e^+ e^-$  sample, the QED background  simulated with GRAPE
is indicated and normalised as described above. 
For the  control distributions of the $\mu^+\mu^-$ sample the non-resonant 
background  is subtracted from the data using
a side band method~\cite{fhuber}. 
This background contains a contribution from non-resonant diffractive  
events, due to pions misidentified 
as muons, in addition to the QED background. 
The data in all distributions are well described by the simulation.
 
\renewcommand{\arraystretch}{1.2} 
\begin{table}
 \begin{center}
   \begin{tabular}{ccccccc}
     \toprule
    Data Set & $E_p$							& 	Process 	& \QSq 							& 	\My							& 	\T									& \Wgp \\
     \midrule
     \multirow{2}{*}{HE}		& 
     \multirow{2}{*}{$\unit[920]{GeV}$}		& 
     elas														& 	 
     \multirow{2}{*}{$< \unit[2.5]{GeV^2}$}	& 	
     $m_p$													&	 
     \multirow{2}{*}{$< \unit[8]{GeV^2}$}	&  
     \multirow{2}{*}{$40 - \unit[110]{GeV}$} \\
    & & pdis			& 	& 	$m_p-\unit[10]{GeV}$			&  \\
    
    \midrule
    \multirow{2}{*}{LE}		& 
     \multirow{2}{*}{$\unit[460]{GeV}$}		& 
     elas														& 	 
     \multirow{2}{*}{$< \unit[2.5]{GeV^2}$}	& 	
     $m_p$													&	 
     \multirow{2}{*}{$< \unit[5^{(\star)}, \, 8]{GeV^2}$}	&  
     \multirow{2}{*}{$25 - \unit[80]{GeV}$} \\
    &	& pdis			& 	& 	$m_p-\unit[10]{GeV}$			&  \\
     \bottomrule
   \end{tabular}
 \end{center}
 \caption{
Kinematic range of the analysis.  
The phase space for elastic and proton-dissociative \jpsi processes is indicated by
elas and pdis, respectively. 
The high- and low-energy data sets are denoted by HE and LE.
${}^{(\star)}$ The phase space restriction is applied only for the 
$\mathrm{d}\sigma / \mathrm{d}t$  cross section measurement.}
 \label{tab:PhaseSpace}
\end{table}

\subsection{Cross section determination and systematic uncertainties}
\label{sec: Cross section determination}
The cross sections are measured for the kinematic ranges as defined in
table~\ref{tab:PhaseSpace}. 
From the unfolded number of events in each signal bin $i$  for the reaction 
$\gamma p \rightarrow \JPSI \rightarrow \ell \ell$,  
the bin-averaged  cross sections are obtained as
\begin{linenomath*}
\begin{align}
\label{eq: Xsection1}
\dfrac{\mathrm{d} \sigma(\gamma p \rightarrow \JPSI)}{\mathrm{d} t} = 
 \dfrac{1}{\Phi_\gamma^T}
 \dfrac{N_{i, t, \ell \ell}}{\mathcal{L}  \cdot \mathcal{B}(\ell \ell) \cdot \Delta t_i} \quad,
\end{align}
\end{linenomath*}
and
\begin{linenomath*}
\begin{align}
\label{eq: Xsection2}
\sigma_{\Wgp}(\gamma p \rightarrow \JPSI) = 
\dfrac{1}{\Phi_\gamma^{T, i, \Wgp}}
\dfrac{N_{i, \Wgp, \ell\ell}}{\mathcal{L}  \cdot \mathcal{B}(\ell \ell) } \quad,
\end{align}
\end{linenomath*}
where the variable $\Phi_\gamma^T$ is the transverse polarised photon flux~\cite{DiffVM},
$\Phi_\gamma^{T, i, \Wgp}$  the transverse polarised photon flux per $\Wgp$ bin,
$\Delta t_i$ the bin width in $t$, 
$\ell \ell=ee$ or $\mu\mu$ depending on the decay channel,
$N_{i, t, \ell \ell}$ and  $N_{i, \Wgp, \ell \ell}$ are
the numbers of unfolded signal
events in the corresponding bins of $t$ or $\Wgp$, $\mathcal{L}$ is the integrated luminosity,  and 
$\mathcal{B}(ee)= 5.94\%$,  $\mathcal{B}(\mu\mu)= 5.93\%$ are
the $\JPSI$ branching fractions~\cite{Beringer:1900zz}. 

The systematic uncertainties on the $\JPSI$ cross section measurement are determined
by implementing shifts due to each source of   uncertainty in the simulation and  propagating
the resulting variations in the  unfolding matrices to the result.
Those uncertainties which are uncorrelated between the two decay modes are classified as individual 
systematic uncertainties, 
while the uncertainties correlated between the $e^+e^-$ and $\mu^+\mu^-$ samples are 
referred to as common systematic uncertainties.

The individual systematic  uncertainties are as follows.
\begin{description}
\item[Lepton identification] 	
			The efficiency of the simulated muon identification is reweighed to agree with that determined 
			from data.  The efficiency was determined with a $\JPsiTomumu$ sample, selected with 
			at least one identified muon. The second muon is then probed to evaluate the single muon
			identification efficiency.
 			The uncertainty on these weights is determined 
 			from the remaining difference between the simulation compared to data~\cite{fhuber}. 
 			The resulting uncertainty
 			on the cross sections is $2\%$ at most.			
			
			The cut value on the electron discriminator $D$ is varied by $\pm 0.04$
 			around its nominal value of $0.8$, which covers the differences in the $D$-distribution 
 			between  simulation and  data. 
 			The uncertainty propagated to the cross section is below $2\%$. 
\item[Signal extraction] 	
			The uncertainty on the number of signal events due to the fitting procedure 
			of the $\Mmumu$ invariant mass distributions is determined by a bias study 
			as described in~\cite{fhuber}  and is typically $\approx 1\%$ but can rise to
			$\approx 5\%$ for the lowest $\Wgp$ bin of the proton-dissociative cross section. 
			
			The uncertainty on the background subtraction procedure for the $e^+e^-$ sample 
			is estimated by determining the background normalisation factor  with data at very low
			invariant di-electron masses $\Mee$ and agreement with the default method is found 
			within $20\%$. The 
			corresponding variation on the 
			background is propagated to the differential cross sections which vary between
			$3\%$ for bins with a low background to $11\%$ for bins with a higher 
			background contribution.

\item[Branching ratio] 	
			The relative uncertainty on the branching ratio  for the muon and electron decay channels is
			$1\%$~\cite{Beringer:1900zz}.	
\end{description}

The following systematic uncertainties are have components contributing to the channel-specific
 individual and the common systematic uncertainties.
\begin{description}

\item[Trigger]
			The trigger efficiency is typically $80 \%$ and 
			is taken from the simulation. The trigger simulation is verified by a comparison
			to data in a sample of $\JPsi$ mesons in deep-inelastic-scattering triggered independently
			on the basis of the scattered beam positron. 
			A small difference of $3 \%$ is observed  between the data and the simulation for 
			$\JPsi$ events decaying into muons. This difference is accounted for by a corresponding 
			upwards shift of the efficiency in the simulation. 
			No such correction is necessary  for electrons.
			The remaining uncertainty is estimated to be  $2\%$ uncorrelated between 
			the $e^+e^-$ and $\mu^+\mu^-$ samples, 
			i.e. treated as individual uncertainties, and  $2\%$ correlated between the two decay channels, i.e.
			treated as a common uncertainty.	
\item[Track finding efficiency] 	
			The uncertainty due to the track reconstruction efficiency in the CTD is estimated to be $1\%$ 
			per track~\cite{mbrinkmann}.
			For electron tracks an additional $1\%$ is applied, to account for the different hit finding
			efficiency  due to bremsstrahlung effects.  
			Since the uncertainty on the track finding efficiency
			affects both selected tracks coherently, a common uncertainty of $2\%$
			is applied to both samples and an additional $2\%$ is applied for the 
			electron sample. 
\end{description}

The following  common systematic  uncertainties  are considered.
\begin{description}
 \item[Tagging] 	
 			The systematic uncertainty arising from   the tagging condition
 			is estimated by varying separately the simulated tagging efficiency
 			for each detector used. The variations cover any possible shift in the individual 
 			relative efficiency distributions, and are $20\%$ for the condition from the forward LAr calorimeter,
			$5\%$ for the PLUG  and $1\%$ for the FTS~\cite{fhuber}.  The resulting uncertainties
 			on the cross sections are typically a few percent, but  reach 
 			$30\%$  at the highest  $|t|$ values of the elastic 
 			$\mathrm{d}\sigma / \mathrm{d}t$ cross section.

\item[Empty calorimeter] 	
		The uncertainty on the cut ensuring an empty calorimeter is obtained by
		varying the maximum allowed $\sum_i E_i^2$ from $2.25\gevsq$ to
		$2.75\gevsq$ in the simulation. 
		This results in an uncertainty of typically $5\%$ for the 
 		proton-dissociative cross sections. For the elastic cross sections this variation is  negligible
 		for most bins, except for the highest bin in $|t|$, where it reaches to $13\%$. 				
 \item[MC modelling] 	
 			The model uncertainty in the MC simulation due to uncertainties in the
 			dependences on $t$, $\Wgp$ and $M_Y$ 
 			is determined by varying the fit parameters of the  weighting procedure  
 			within the errors obtained in a dedicated fit
			of the forward energy flow~\cite{fhuber}. For the 
 			cross section as a function of $\Wgp$ the corresponding uncertainties are below $4\%$, whereas
 			for the cross sections differential in $t$, values around $10\%$
 			are obtained for the high $|t|$ bins.	
\item[Luminosity]
		The integrated luminosity is known  to within $\pm 2.7\%$ for the HE data set and  
		to within $\pm4\%$ for the  LE data set~\cite{QEDComptonAnalysis}.
\item[$\boldsymbol{\psi(2S)}$ background]
		Background  from $\psi(2S)$ decays to $\JPsi \, X$  is estimated to contribute  $4\%$ 
		to the selected $\JPsi$ events, and is subtracted from the data prior
		to the unfolding procedure~\cite{Adloff:2002re}. The cross section measurements
		are affected by an uncertainty of $1.5\%$.
\item[$\boldsymbol{\qsq}$ dependence]		
		The $\qsq$ dependence of the cross section is parametrised as 
		$\sigma_{\gamma p} \propto \left( m_\psi^2 + Q^2 \right)^{-n}$ \cite{Aktas:2005xu}.
		The corresponding systematic uncertainty is obtained by varying 
		the parameter $n$ in the range $2.50 \pm 0.09$.
		The cross sections are affected by less than $1\%$. 
\end{description}

The differential cross sections obtained from the electron and the muon data 
agree within uncertainties. The two measurements are combined by taking
into account their individual uncertainties. 
This combination procedure involves the
numerical minimisation of
a standard $\chi^2$ function 
including the full statistical error matrix and the correlated systematic errors 
with nuisance parameters, similar to that defined in~\cite{Aaron:2009bp, Aaron:2009aa}.
All individual uncertainties are incorporated within
this procedure, whereas the common uncertainties are considered after the combination only.
The consistency of the data sets can be verified by looking at the resulting nuisance
 parameters.
None of the nuisance parameters shifts by more than one
standard deviation. 

\begin{figure}[hhh]
\centering
\setlength{\unitlength}{7.5cm}
\includegraphics[width=7.5cm] {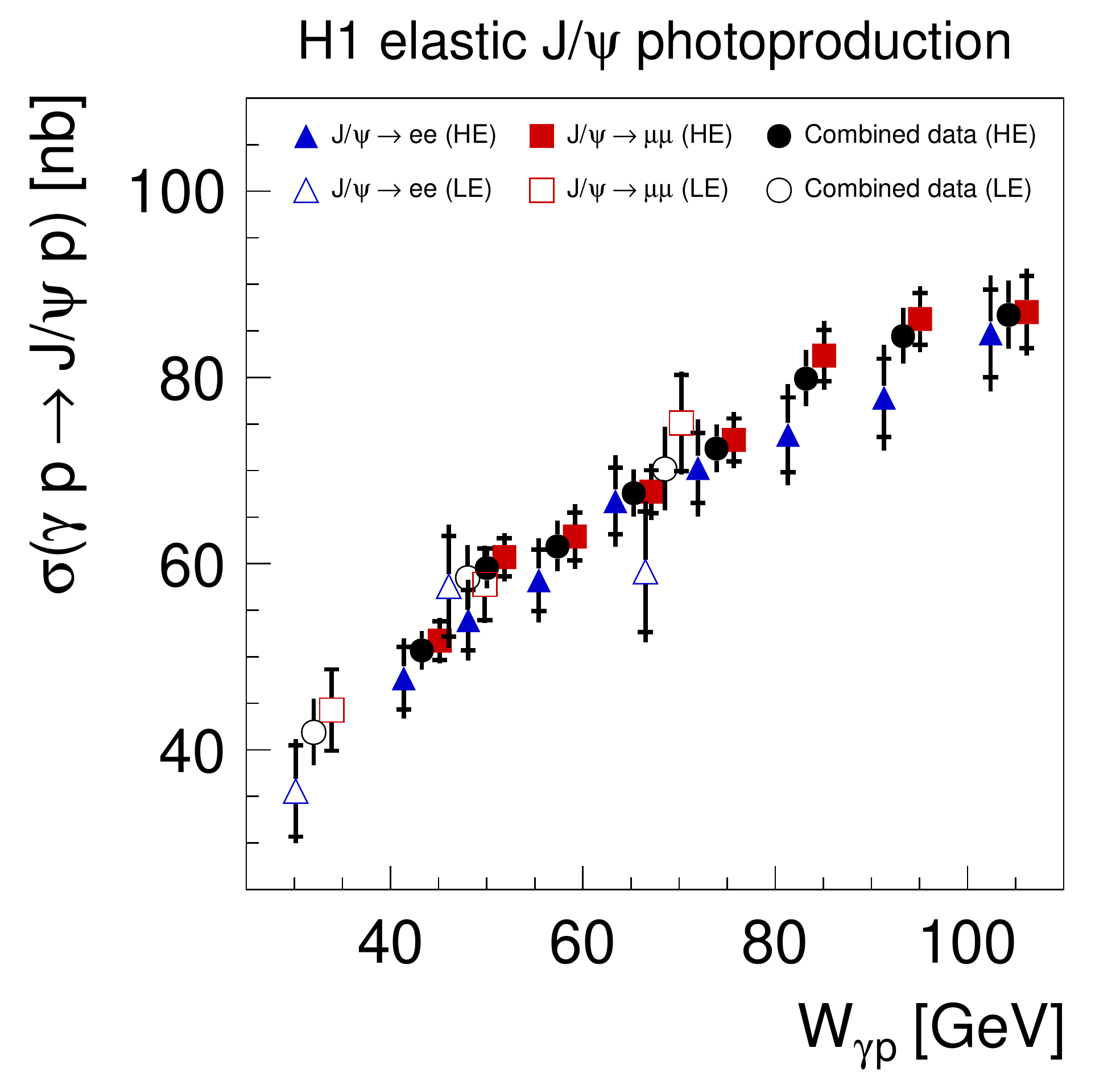}
\put(-0.95, 0.95) {\bf{a)}}
\includegraphics[width=7.5cm] {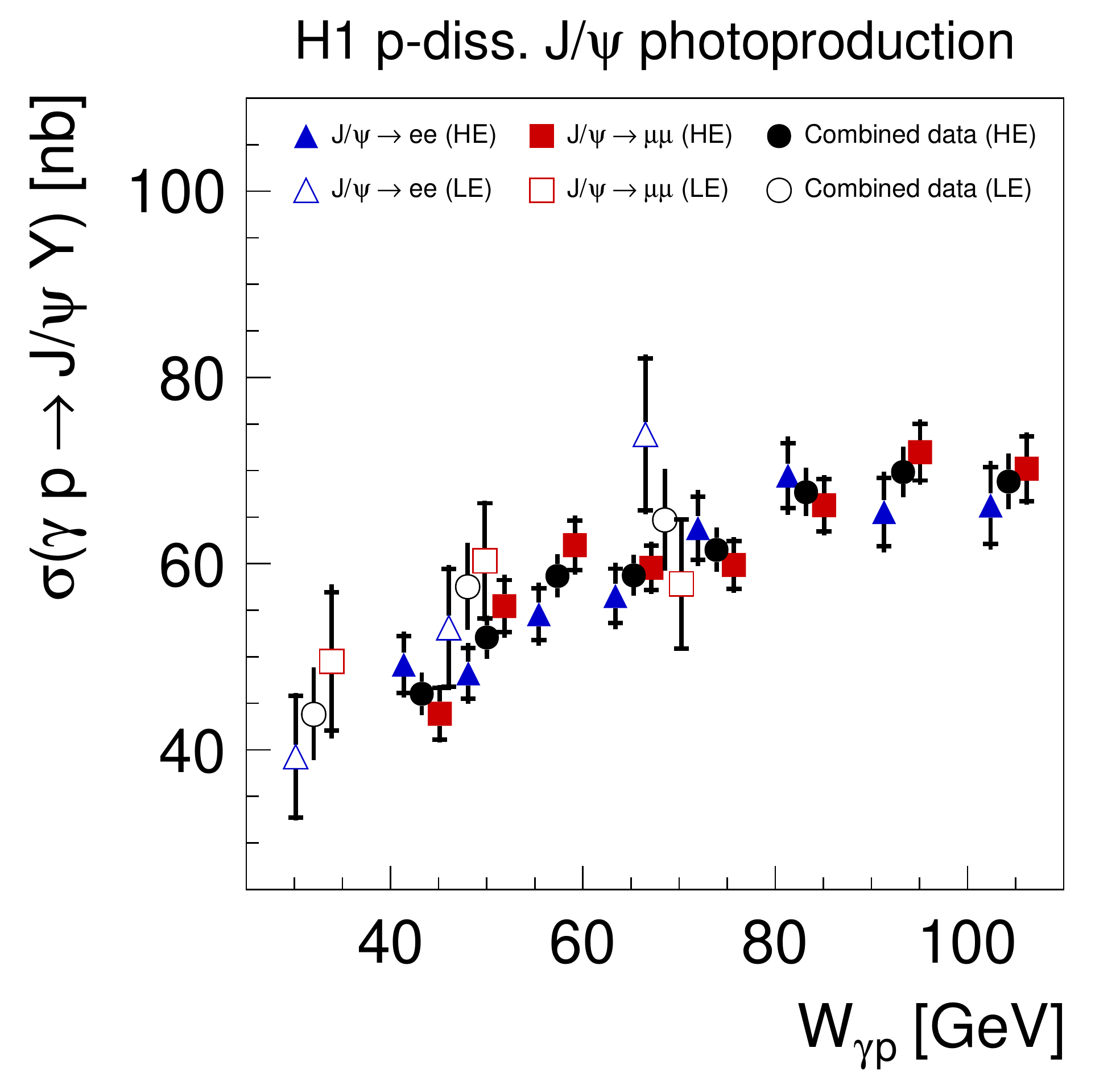}
\put(-0.95, 0.95) {\bf{b)}} \\

\vspace{1cm}
\caption{Combined elastic and proton-dissociative cross sections as a function of $\Wgp$
(circles) compared to the input data from 
$\JPSI\rightarrow e^+e^-$ (triangles) and  $\JPSI\rightarrow\mu^+\mu^-$ 
(squares) 
of the HE and LE data sets. The error bars of the input data indicate the  uncertainty 
composed of the statistical errors (inner error bars) and statistical errors combined with 
all individual systematic uncertainties (full error bars).
The error bars of the combined data points reflect the uncertainty after the combination.
The combined data points are drawn at their bin centres.
The electron and muon data points are shifted in $\Wgp$ for  better visibility.}
\label{fig: DataCombination} 
\end{figure}

Figure~\ref{fig: DataCombination} shows the result of the combination for 
the elastic and proton-dissociative 
cross sections as a function of  $\Wgp$.
The input data obtained in the electron and muon decay channels
    are shown together with the combined data.

\section{Results}
\label{sec: Results}
The elastic and proton-dissociative differential $\JPsi$ cross sections as functions of  $t$ and $\Wgp$ 
are measured in the kinematic ranges defined in table~\ref{tab:PhaseSpace} using 
the decay channels $\JPsiTomumu$ and\linebreak $\JPsiToee$.

Tables~\ref{tab: CS_W}, \ref{tab: CS_HER_t} and~\ref{tab: CS_LER_t} list the combined data points for 
all cross sections together with their uncertainties
and all common systematic uncertainties. 
The input data to the combination procedure, including all individual systematic uncertainties
together with the full covariance matrices of the
combined results can be found in~\cite{webmaterial}.

\subsection{$\boldsymbol{t}$ dependence}
\label{sec: t dependence}
\begin{figure}[hhh]
\centering
\setlength{\unitlength}{7.5cm}
\includegraphics[width=7.5cm] {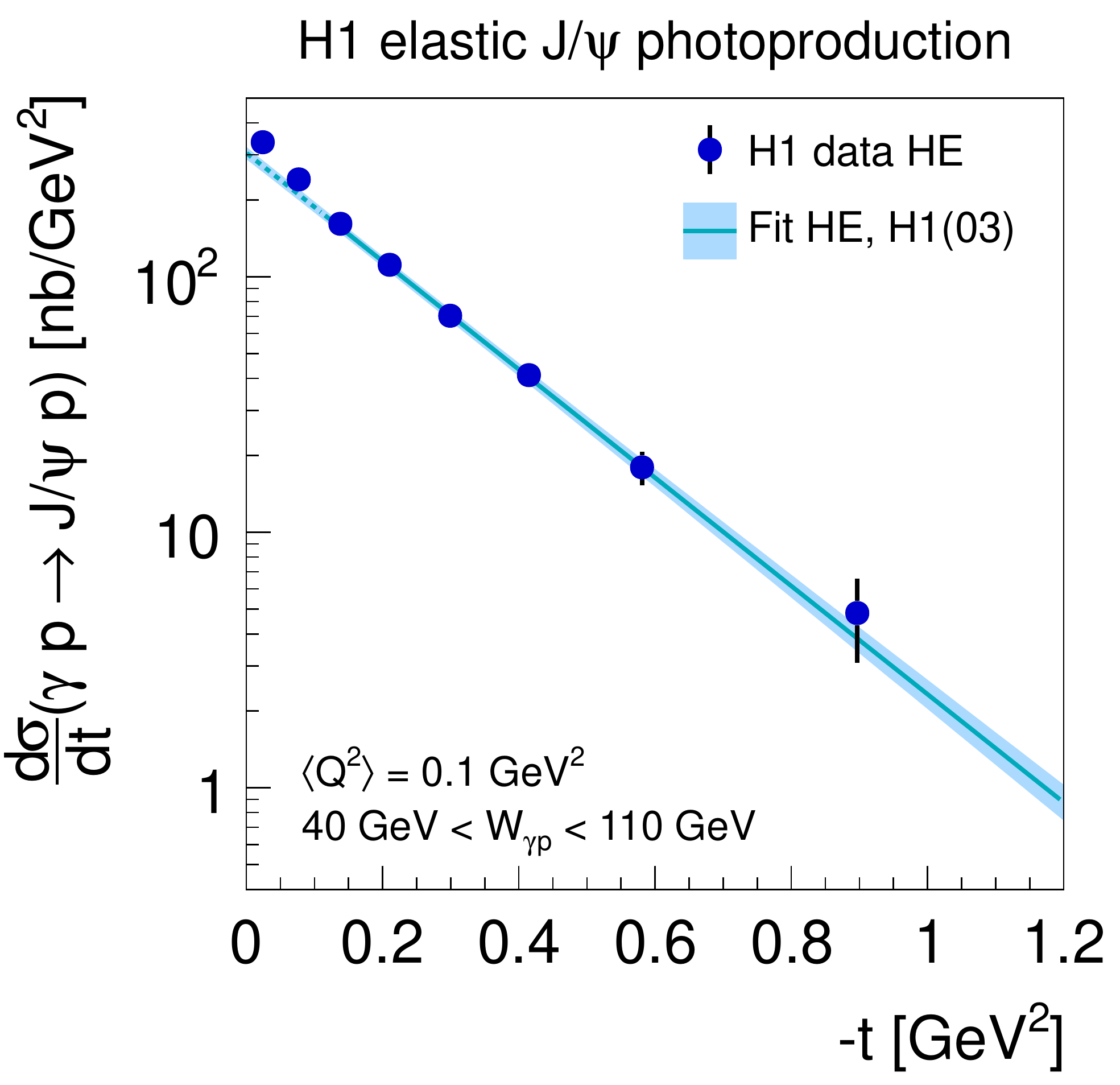}
\put(-0.95, 0.95) {\bf{a)}}
\includegraphics[width=7.5cm] {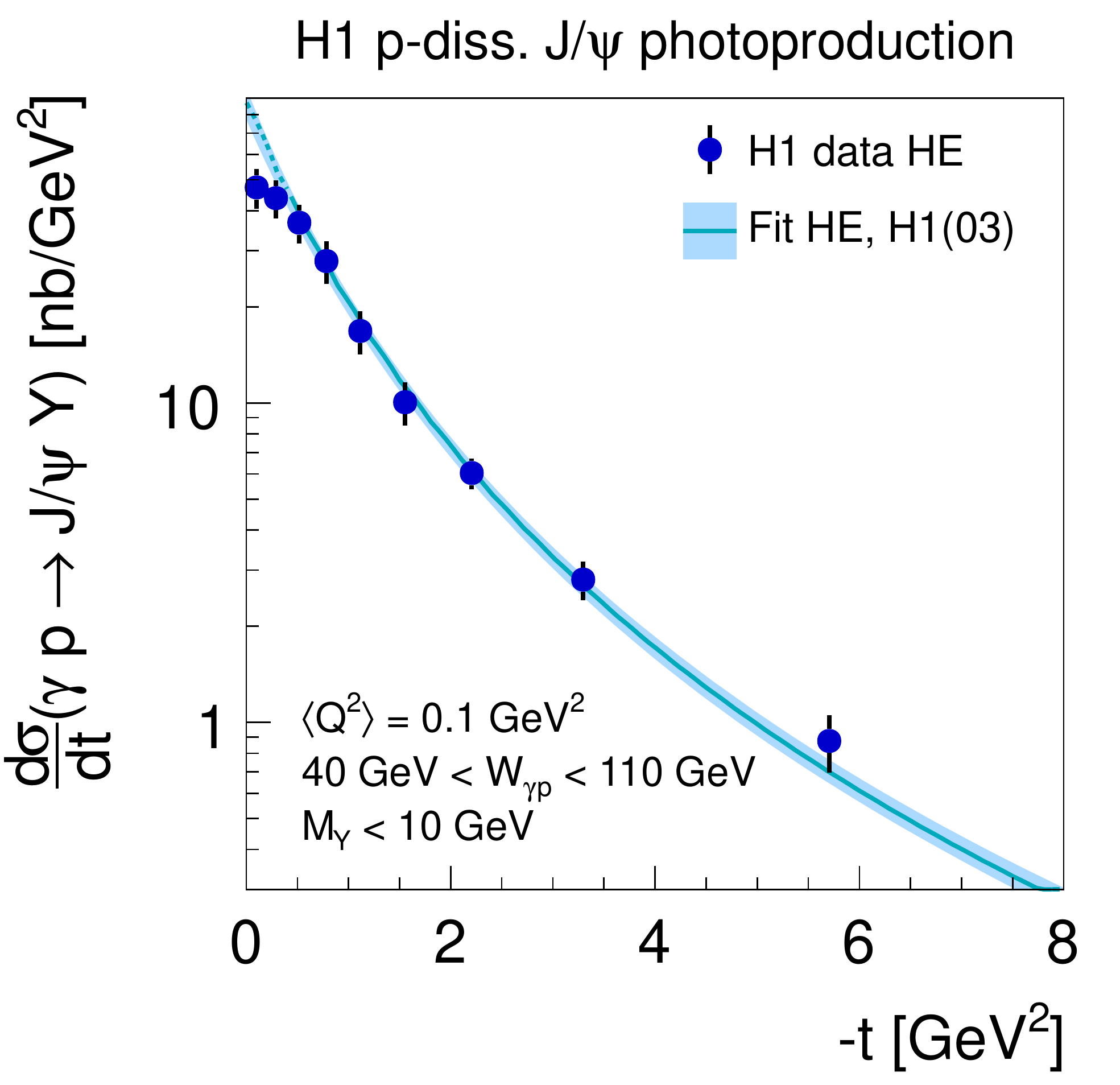}
\put(-0.95, 0.95) {\bf{b)}} \\

\includegraphics[width=7.5cm] {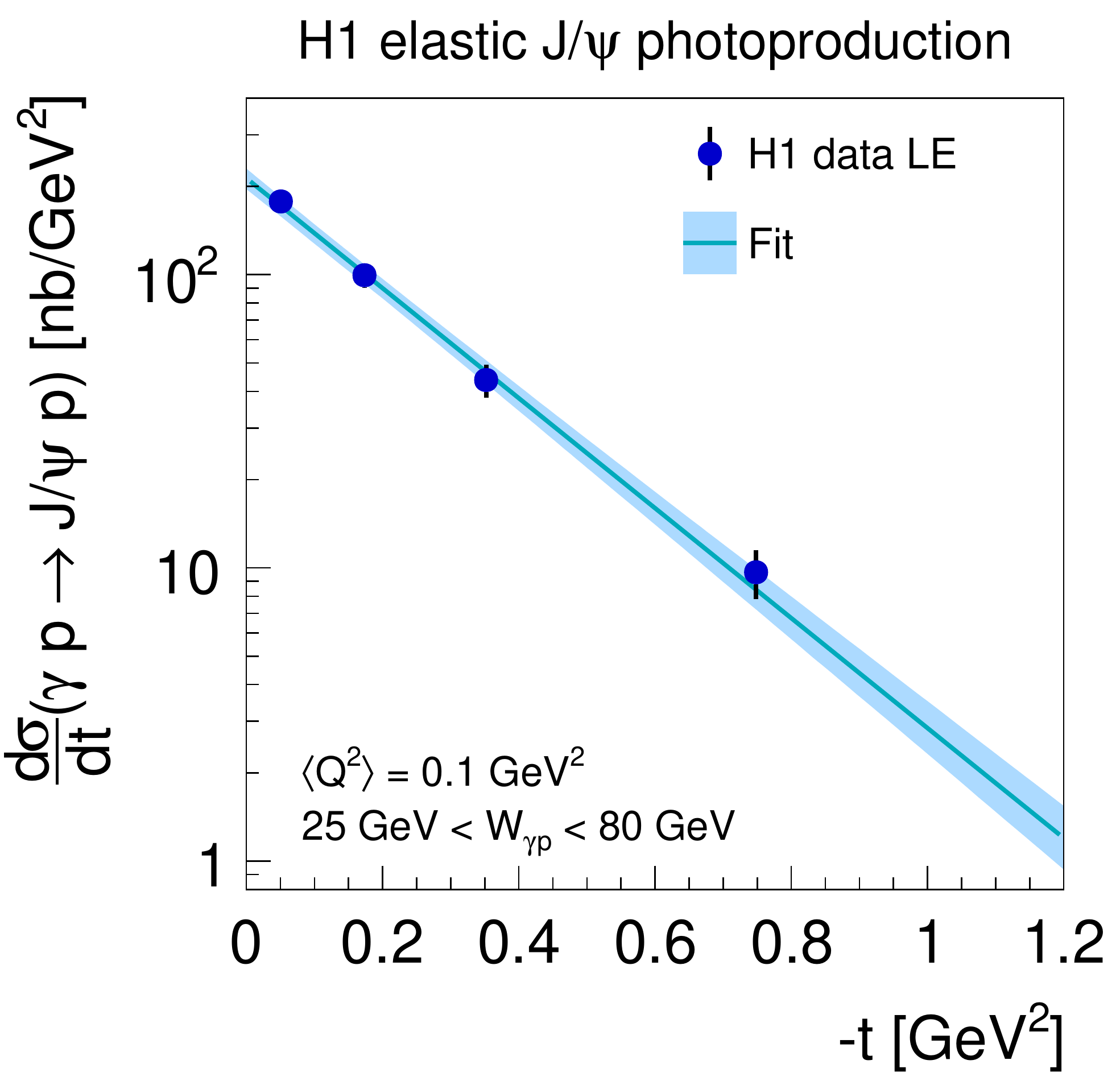}
\put(-0.95, 0.95) {\bf{c)}}
\includegraphics[width=7.5cm] {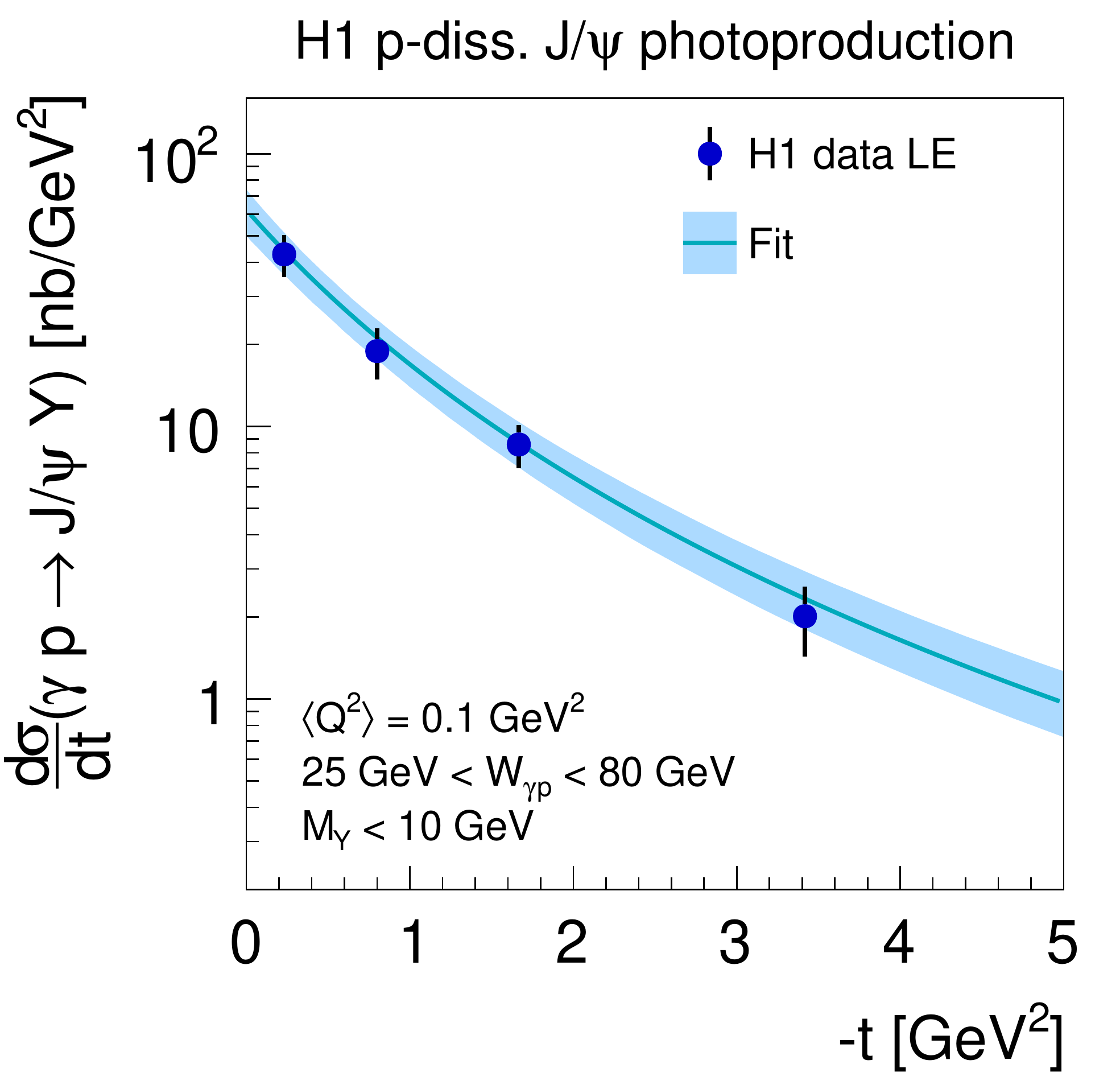}
\put(-0.95, 0.95) {\bf{d)}} \\

\vspace{1cm}
\caption{
Differential $\JPSI$ photoproduction cross sections $\mathrm{d}\sigma / \mathrm{d} t $ 
as a function of the negative squared four-momentum transfer at the
proton vertex, $-t$, as obtained in the high-energy data set for the (a) elastic 
regime and the (b) proton-dissociative regime and as obtained for the
low-energy data set shown in (c) and (d).
The error bars represent  the total errors. Also shown by the curves is a simultaneous fit to 
this measurement and~\cite{Aktas:2003zi}
of the form
$\mathrm{d}\sigma / \mathrm{d}t = N_{el} e^{-b_{el} |t|}$  for the elastic cross sections  and
	$\mathrm{d}\sigma / \mathrm{d}t = N_{pd} (1+(b_{pd}/n) |t|)^{-n}$ 	
 for 
 the proton-dissociative cross sections. 
The fit uncertainty is represented by the spread of the shaded bands.
}
\label{fig: Xsetion_t} 
\end{figure}

Figure~\ref{fig: Xsetion_t} shows the measured elastic and proton-dissociative 
cross sections differential in $-t$, separately for the LE and HE data sets. 
The cross sections fall steeply with increasing $-t$, and shows a clear difference between the shapes 
of  the proton-dissociative  and
elastic distributions.
The proton-dissociative cross section levels off for very 
low values of $|t|$. There is a phase space effect such that for small
$|t|$ it is not possible to produce large masses of $M_Y$. 

\begin{figure}[hhh]
\centering
\includegraphics[width=10cm] {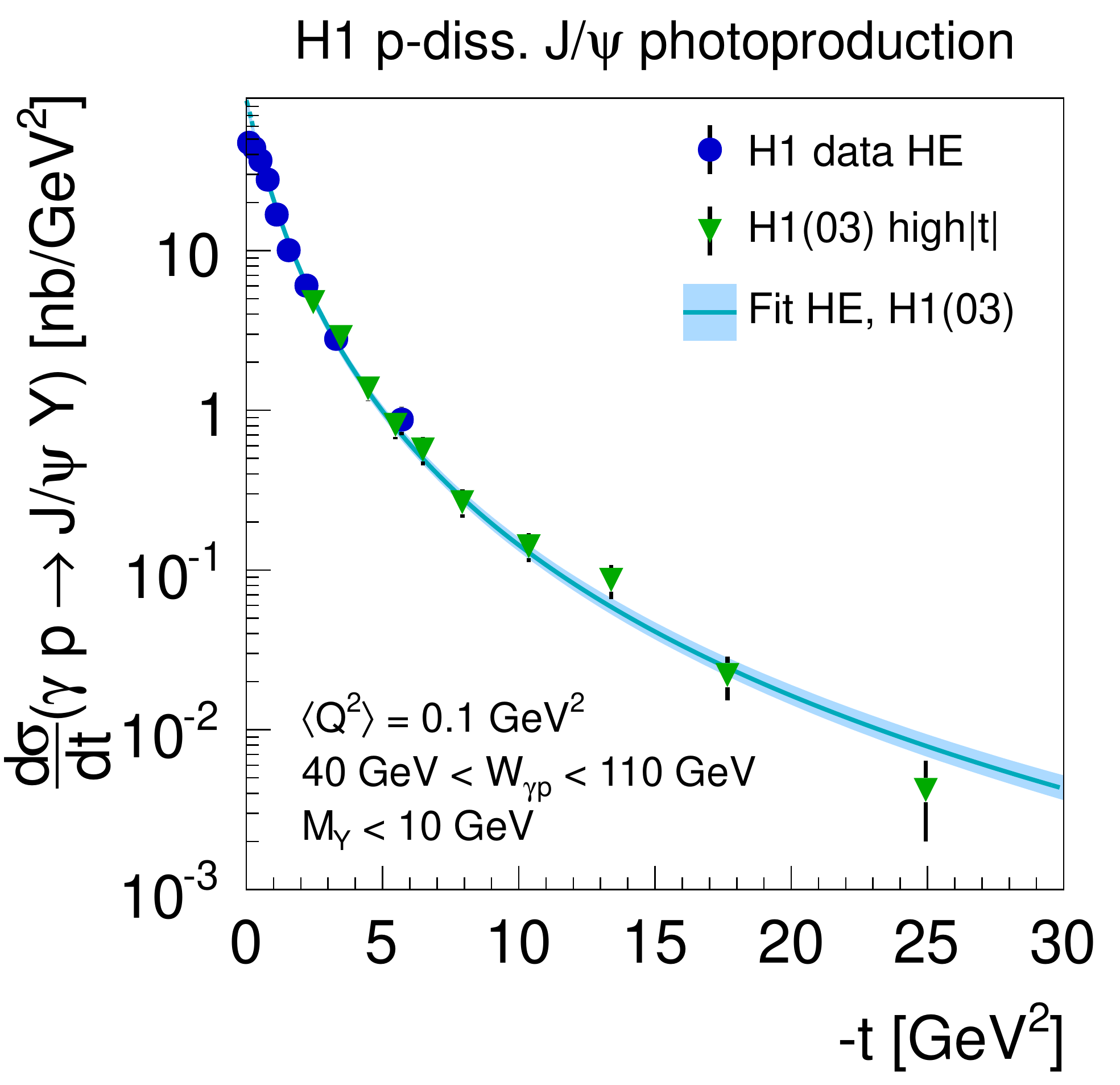}

\vspace{1cm}
\caption{Proton-dissociative cross section as a function of $-t$ (full circles)
compared to a previous measurement at high~$|t|$~\cite{Aktas:2003zi} (triangles)
interpolated to match the $\Wgp$, $\qsq$ and $M_Y$ ranges of the current
measurement. The
curve represents a simultaneous fit to both data sets, the spread of the shaded
band its uncertainty.}
\label{fig: OldMeasurementsPdis} 
\end{figure}

In figure~\ref{fig: OldMeasurementsPdis} the proton-dissociative measurement from
the HE data set as a function of $-t$  is compared to a 
previous analysis~\cite{Aktas:2003zi} covering the region of high $|t|$, which
is completely dominated by proton-dissociative
events.
The high~$|t|$ data~\cite{Aktas:2003zi} are adjusted to the $\Wgp$, $\qsq$ and $M_Y$
ranges of the present analysis by applying a phase space correction
of about $7\%$. Comparing the two measurements, 
the present proton-dissociative cross sections extend the
reach to small values of $|t|$.
In the overlap region  $2 < |t| < 8 \gevsq$ the two measurements
agree.

The  elastic and proton-dissociative differential cross sections $\mathrm{d} \sigma / \mathrm{d}t$
are  fitted simultaneously, using a $\chi^2$-function~\cite{Aaron:2009bp, Aaron:2009aa}
based on the error matrix obtained in the combination procedure 
and all common systematic uncertainties. 
The elastic cross section is parametrised as
$\mathrm{d}\sigma / \mathrm{d}t = N_{el}\, e^{-b_{el} |t|}$.
For the proton-dissociative cross section 
$\mathrm{d}\sigma / \mathrm{d}t = N_{pd}\,  (1+(b_{pd}/n) |t|)^{-n}$ is chosen,
which interpolates between an exponential at low $|t|$ and a power law behaviour at high values of $|t|$.
The fits are performed separately for the HE and the LE measurements. In the case
of the HE data the previously measured high $|t|$ data are included in the fit. This fit
yields a value of $\chi^2/ \mathrm{NDF} = 26.6 / 18$ after excluding the two lowest 
$t$ data points in both the elastic and the proton-dissociative channel. 
For fit of the LE data set, the parameter $n$ is fixed to the value obtained from the HE data set, 
since the LE data are not precise enough to constrain
$b_{pd}$ and $n$ simultaneously.
The obtained parametrisations for the elastic and proton-dissociative cross sections
are compared to the data in figure~\ref{fig: Xsetion_t} and figure~\ref{fig: OldMeasurementsPdis}.
Table~\ref{tab:CSFitResults_t} summarises the fit parameters and their uncertainties.

The elastic cross section data for $-t > 0.1\gev $~are well described by the exponential parametrisation.
They fall much faster with increasing $|t|$ 
than the proton-dissociative cross section even at
    small $|t|$, which is reflected in 
the values for $b_{el} $ and $b_{pd}$. The  value extracted for
$b_{el}$ is compatible with previous results~\cite{Aktas:2005xu},
although the previous fit was done as a function of $p^2_{T, \JPSI }$ rather than $-t$. 
Some difference between the $b_{el}$ values for the LE and HE data is expected~\cite{Aktas:2005xu} due to the different
ranges  in $\Wgp$ corresponding to  $\langle \Wgp \rangle = 78 \gev$ for the HE data set
and $\langle \Wgp \rangle = 55 \gev$ for the LE data. 

\subsection{Energy dependence}
\label{sec: W dependence}

\begin{figure}[hhh]
\centering
\setlength{\unitlength}{7.5cm}
\includegraphics[width=7.5cm] {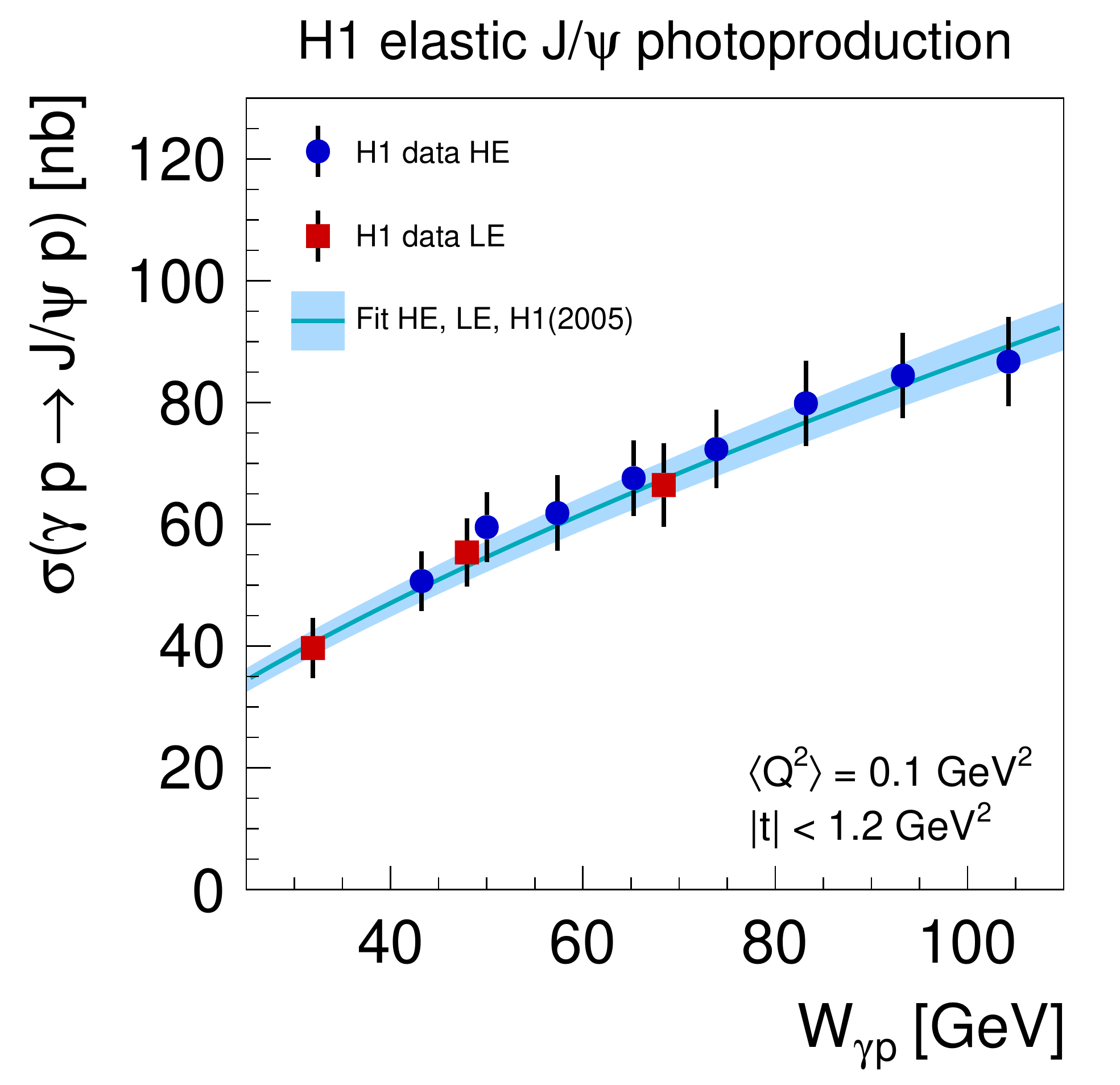}
\put(-0.95, 0.95) {\bf{a)}}
\includegraphics[width=7.5cm] {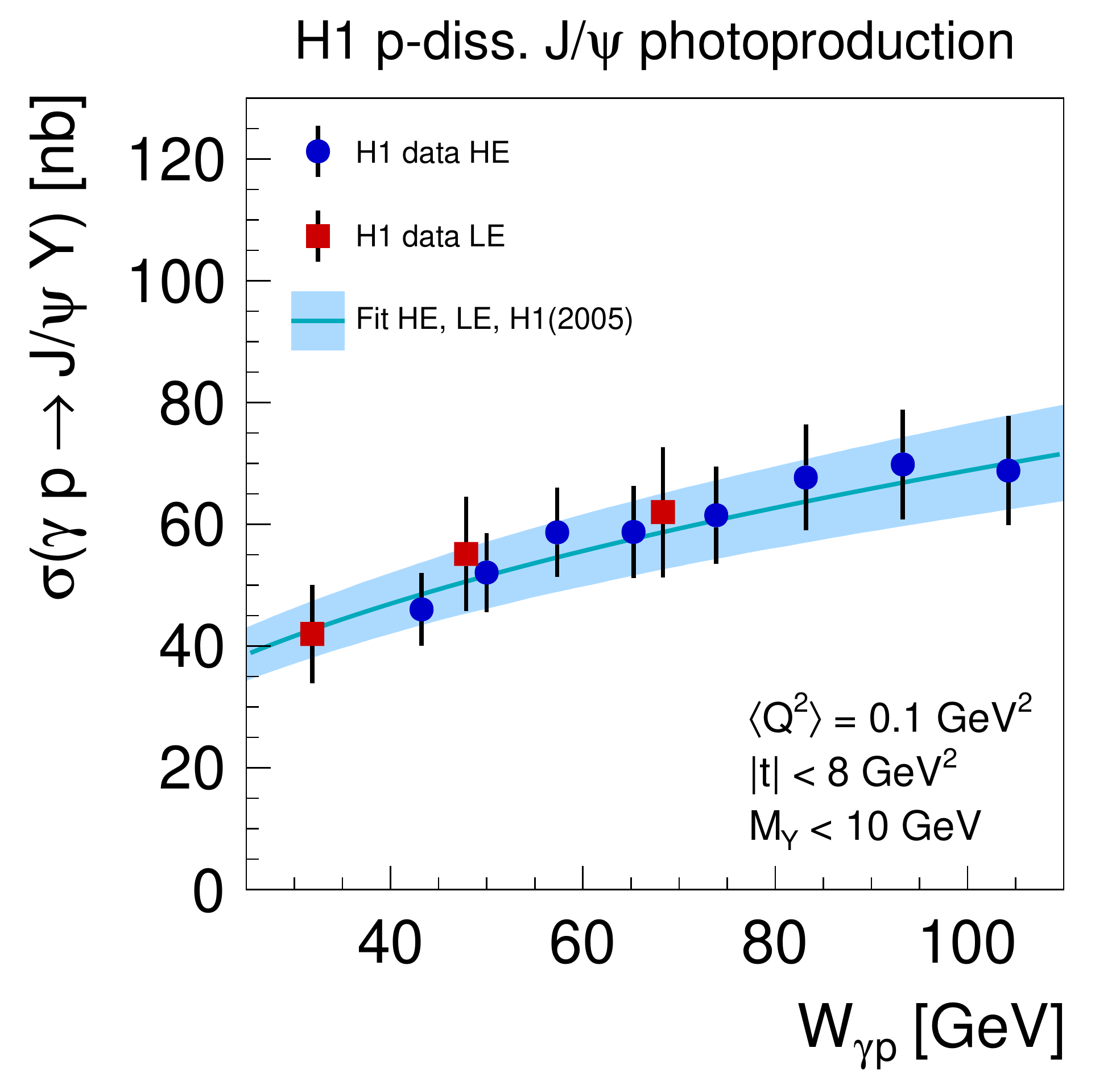}
\put(-0.95, 0.95) {\bf{b)}}\\
\includegraphics[width=7.5cm] {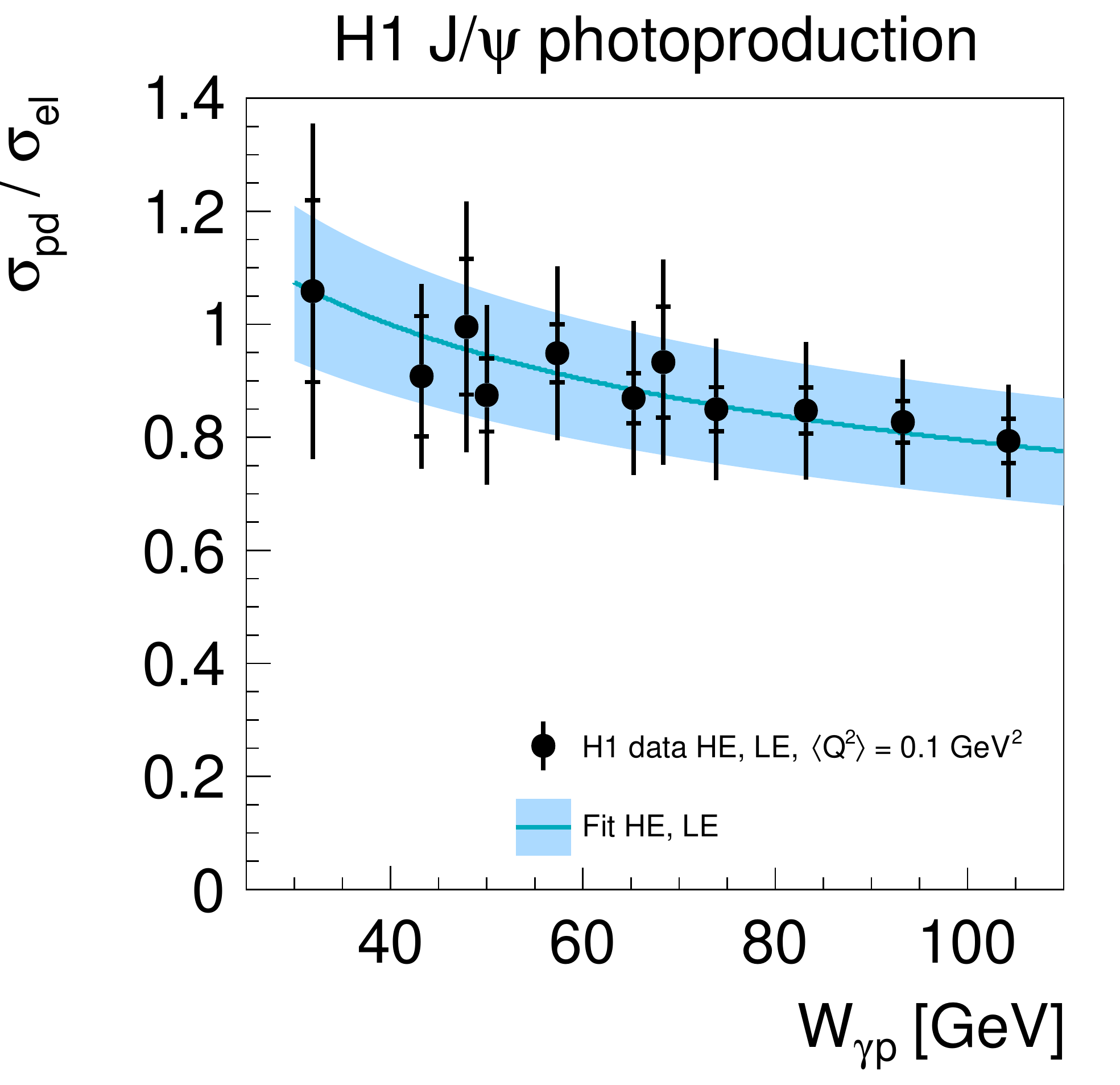}
\put(-0.95, 0.95) {\bf{c)}}

\vspace{1cm}
\caption{$\JPSI$ photoproduction cross sections  
as a function of the photon proton centre-of-mass energy $\Wgp$ 
for (a) the elastic and (b) the proton-dissociative regime.
The data from the high-energy  data set are shown by circles,
the data from the low-energy data set as  squares.
The error bars represent the total errors. Shown by the curves is the simultaneous fit to the data from
this measurement and~\cite{Aktas:2005xu}, see figure~\ref{fig: OldMeasurementsElas}.
The fit uncertainty is represented by 
the shaded bands.
In (c) 
the ratio of the proton-dissociative to elastic $\JPSI$ photoproduction cross section  
is shown. 
The  data are presented as full circles and the vertical
bars indicate the total uncertainties, including normalisation
uncertainties. The inner error bars represent
the bin-to-bin uncorrelated errors, determined in an approximative procedure.
The curve is the ratio of the fits 
shown in (a) and (b). 
The shaded band indicates the uncertainty on the ratio obtained from the fit uncertainties.
}
\label{fig: Xsetion_W} 
\end{figure}

The measured elastic and proton-dissociative cross sections as a function of $\Wgp$ are 
shown in figures~\ref{fig: Xsetion_W}. The elastic and
proton-dissociative cross sections are of similar size at the lowest $\Wgp=30 \, \gev$
accessed in this analysis. The elastic cross
section rises faster with increasing $\Wgp$ than the proton-dissociative one. 
The ratio of the proton-dissociative to the elastic cross section  as a function of
$\Wgp$ is also shown in figure~\ref{fig: Xsetion_W}. 
The ratio decreases from $1$ to $0.8$ as $\Wgp$ increases from $30\, \gev$ to $100 \, \gev$.
When calculating the ratio no attempt is made to extrapolate the elastic
measurement to $-t = 8 \gevsq$. The corresponding correction is estimated
to be smaller than $1\%$. 
\begin{table}[t]
 \centering
 \begin{tabular}{llcll}
   \toprule
   Data period & Process & Parameter & Fit value& Correlation\\
   \midrule
   HE 	& \elas 	& $b_{el}$	& $\unit[(4.88 \pm 0.15)]{GeV^{-2}}$ & 
   \begin{scriptsize}
  \begin{tabular}{ l @{=}r } 
  $\rho(b_{el}, N_{el})$& 0.50 \\
  $\rho(b_{el}, b_{pd})$& 0.49 \\
  $\rho(b_{el}, n)$& -0.21 \\
  $\rho(b_{el}, N_{pd})$& 0.68 
      \vspace*{2mm}\\
   \end{tabular} 
  \end{scriptsize}  \\
   			& 				& $N_{el}$	& $\unit[(305 \pm 17)]{nb / GeV^{2}}$ &
      \begin{scriptsize}
  \begin{tabular}{ l @{=}r } 
  $\rho(N_{el}, b_{pd})$& 0.23 \\
  $\rho(N_{el}, n)$& -0.07 \\
    $\rho(N_{el}, N_{pd})$& 0.46 
          \vspace*{2mm}\\
   \end{tabular} 
  \end{scriptsize}  \\			
       		& \pdis 	& $b_{pd}$ 	& $\unit[(1.79 \pm 0.12)]{GeV^{-2}}$&
       \begin{scriptsize}
  \begin{tabular}{ l @{=}r } 
  $\rho(b_{pd}, n)$& -0.78 \\
  $\rho(b_{pd}, N_{pd})$& 0.76 
        \vspace*{2mm}\\
   \end{tabular} 
  \end{scriptsize}  \\	
       		&       		& $n$ 	&  $3.58 \pm 0.15$&
  \begin{scriptsize}
  \begin{tabular}{ l @{=}r } 
  $\rho(n, N_{pd})$& -0.46 
        \vspace*{2mm}\\
   \end{tabular} 
  \end{scriptsize}  \\
   			& 				& $N_{pd}$	& $\unit[(87 \pm 10)]{nb/ GeV^{2}}$ &\\ 
   \midrule
   LE 	& \elas 	& $b_{el}$ 	& $\unit[(4.3\pm 0.2)]{GeV^{-2}}$&
   \begin{scriptsize}
  \begin{tabular}{ l @{=}r } 
  $\rho(b_{el}, N_{el})$& 0.37 \\
  $\rho(b_{el}, b_{pd})$& 0.10 \\
  $\rho(b_{el}, N_{pd})$& 0.41 
      \vspace*{2mm}\\
   \end{tabular} 
  \end{scriptsize}  \\
    		& 				& $N_{el}$	& $\unit[(213 \pm 18)]{nb/ GeV^{2}}$ &
 \begin{scriptsize}
  \begin{tabular}{ l @{=}r } 
  $\rho(N_{el}, b_{pd})$& -0.24 \\
    $\rho(N_{el}, N_{pd})$& -0.10 
          \vspace*{2mm}\\
   \end{tabular} 
  \end{scriptsize}  \\			
       		& \pdis 	& $b_{pd}$ 	& $\unit[(1.6\pm 0.2)]{GeV^{-2}}$&
        \begin{scriptsize}
  \begin{tabular}{ l @{=}r } 
  $\rho(b_{pd}, N_{pd})$& 0.53 
        \vspace*{2mm}\\
   \end{tabular} 
  \end{scriptsize}  \\	
       		&       		& $n$ 	& $3.58\: \mathrm{(fixed\: value)}$& \\
    		& 				& $N_{pd}$	& $\unit[(62 \pm 12)]{nb/ GeV^{2}}$ &\\ 
   \bottomrule
 \end{tabular}
 \caption{
	Parameter values obtained from the   fits to the differential cross sections $\mathrm{d}\sigma / \mathrm{d}t$,
	including their errors and correlations. 
 	The fit functions are described in the text.
 	HE and LE denote the high- and low-energy
   	data sets, respectively. }
 \label{tab:CSFitResults_t}
\end{table}

\begin{figure}[hhh]
\centering

\includegraphics[width=10cm] {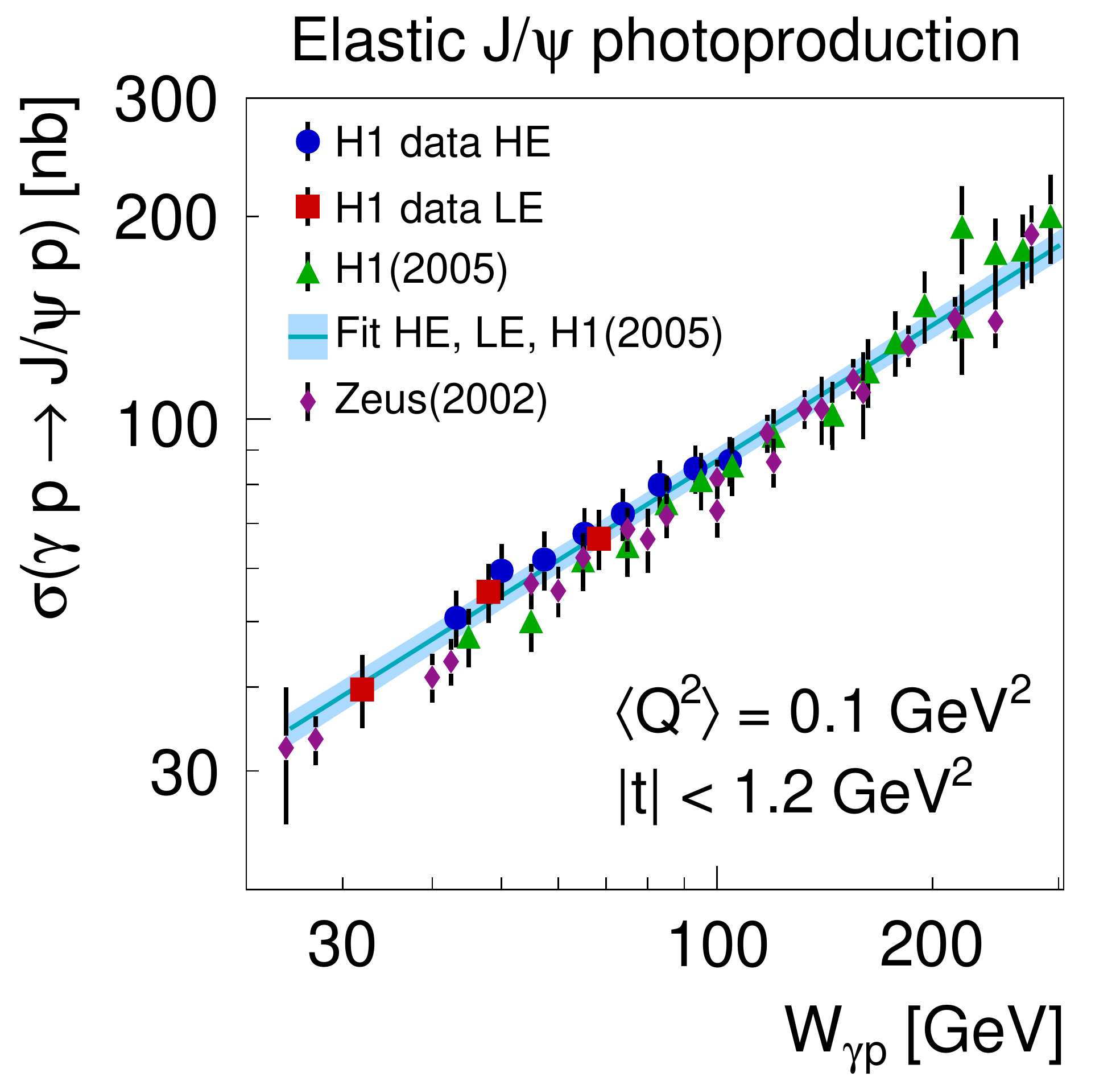}

\vspace{1cm}
\caption{Elastic cross sections as a  function of $\Wgp$ from this measurement
compared to previous measurements at HERA~\cite{Aktas:2005xu, Chekanov:2002xi}. The
shaded band represents a fit to the present data  and~\cite{Aktas:2005xu} together with its uncertainties.
}
\label{fig: OldMeasurementsElas} 
\end{figure}

In figure~\ref{fig: OldMeasurementsElas} the elastic cross section measurements
of this analysis  are compared to previous measurements at HERA~\cite{Aktas:2005xu,Chekanov:2002xi}.
The LE data extend the range accessible  in $\Wgp$ to lower values when
compared to previous H1 measurements~\cite{Aktas:2005xu}.
The HE data have a large overlap with previous H1
measurements in the region $40 \gev < \Wgp < 110 \gev$ and  show a similar
precision. 
Within normalisation uncertainties, the previous
measurements and the new data are in agreement.

The  measured elastic and proton-dissociative cross sections as a function of $\Wgp$, shown
in figure~\ref{fig: Xsetion_W}, are fitted simultaneously, taking into account the 
correlations between the proton-dissociative and the elastic cross sections. 
The fit also includes data from a previous measurement~\cite{Aktas:2005xu} shown
in figure~\ref{fig: Xsetion_W},  with a  normalisation uncertainty
of  $5\%$ and all other systematic uncertainties treated as uncorrelated. 
As  parametrisation two power law functions of the form
$\sigma = N \left( \Wgp / W_{\gamma p, 0}  \right)^ \delta$ 
with $W_{\gamma p, 0}  = 90 \gev$ are used with separate sets of
parameters for the elastic and the proton-dissociative cases.
The $\chi^2$-function is defined in the same manner as for fits
of the $t$-dependences. 

The result of the fit is compared to the measurements in figures~\ref{fig: Xsetion_W}
and in figures~\ref{fig: OldMeasurementsElas}. 
The parametrisation describes the data well ($\chi^2/ \mathrm{NDF} = 32.6 / 36$). 
The fitted parameters  are given in table~\ref{tab:CSFitResults_W} together with their
uncertainties and correlations. 
In Regge phenomenology the parameter $\delta$ can be related to the pomeron
trajectory $\alpha(t) = \alpha(0) + \alpha' \cdot t$ by 
$\delta(t) = 4 (\alpha(t) -1)$.
Using the values $\alpha'_{el} = 0.164 \pm 0.028 \pm 0.030 \gevsqInv$~\cite{Aktas:2005xu} and 
$\alpha'_{pd} = -0.0135 \pm 0.0074 \pm 0.0051 \gevsqInv$~\cite{Aktas:2003zi}, 
together with the mean values of $t$ for the elastic and proton-dissociative 
measurements, $\langle t \rangle = -0.2 \gevsq$ 
and $\langle t \rangle = -1.1 \gevsq$, one
can estimate $\alpha(0)$ for the elastic and proton-dissociative process from these measured parameters.
The obtained values of $\alpha(0)_{el} = 1.20 \pm 0.01$ and $\alpha(0)_{pd} = 1.09 \pm 0.02$
are in agreement with the results from~\cite{Aktas:2005xu, Chekanov:2002xi, Aaron:2009xp}.

The direct comparison between $\delta_{el}$ and $\delta_{pd}$ is made 
by looking at the ratio of the two cross sections, shown
in figure~\ref{fig: Xsetion_W}. The ratio is parametrised as
$N_R \left( \Wgp / W_{\gamma p, 0}  \right)^ {\delta_R}$ with $W_{\gamma p, 0}  = 90 \gev$, 
$N_R = N_{pd}/N_{el} = 0.81 \pm 0.10$ and
$\delta_R = \delta_{pd} - \delta_{el} = -0.25 \pm 0.06$, taking all correlations into account. 
Qualitatively the decrease of this ratio with increasing $\Wgp$  has been  predicted 
in~\cite{Gotsman:2007pn} as a consequence of the non-unit and $\Wgp$ dependant
survival probability for the proton dissociation process.

\begin{figure}[hhh]
\centering
\includegraphics[width=15cm] {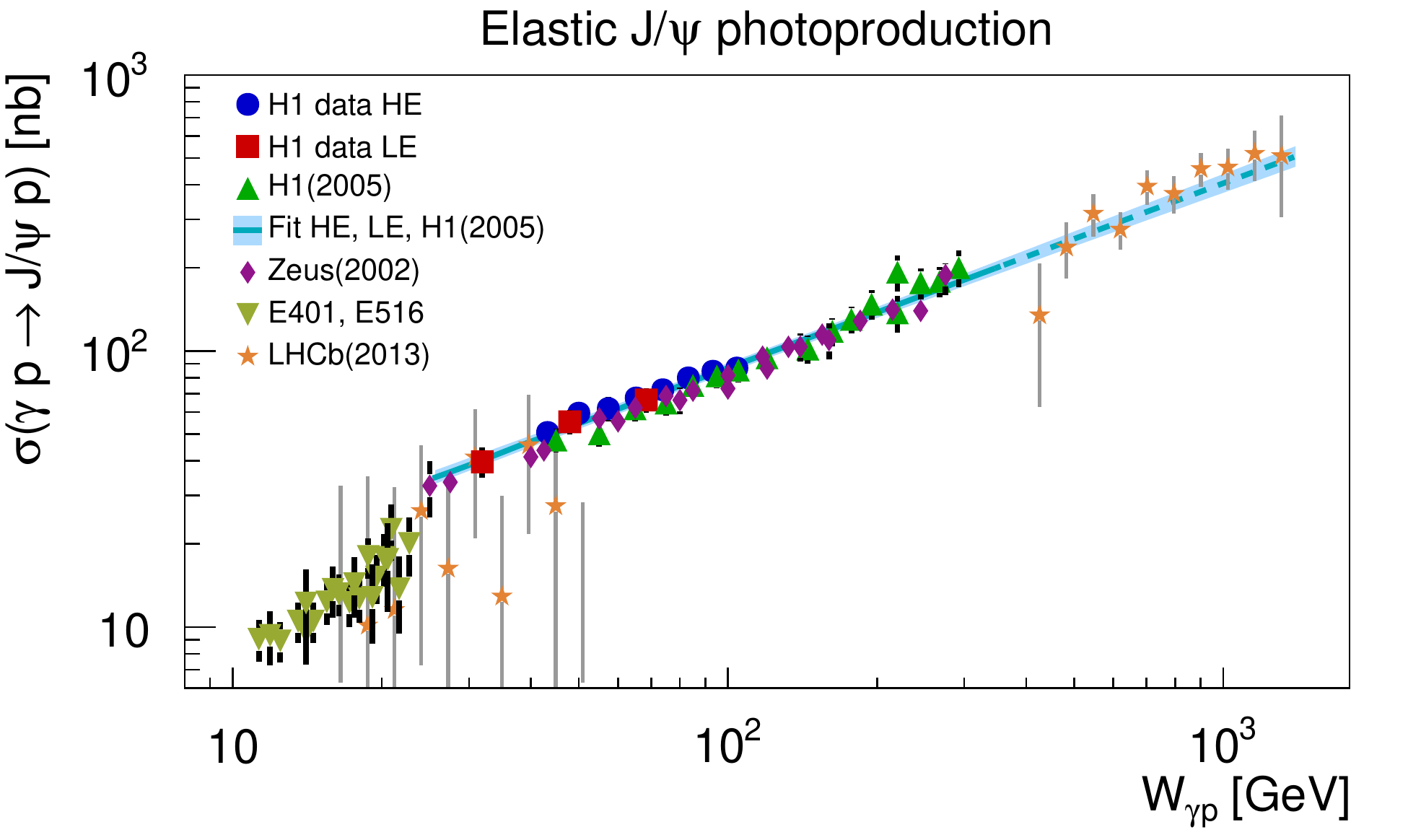}

\vspace{1cm}
\caption{
Compilation of 
elastic $\JPsi$ production cross section measurements including
this measurement, previous 
HERA results~\cite{Aktas:2005xu, Chekanov:2002xi}, 
results from fixed target experiments~\cite{Binkley:1981kv, Denby:1983az}
 and from LHCb~\cite{Aaij:2013jxj}. 
 Also presented is the fit to the H1 data only, indicated
by the curve. The fit is extrapolated in $\Wgp$ from the range of the input data to
higher values, as shown by  the dashed curve.
The shaded band indicates the uncertainty on the fit.
}
\label{fig: Xsetion_WSummary_FixedTarged_LHCb} 
\end{figure}

In figure~\ref{fig: Xsetion_WSummary_FixedTarged_LHCb} a compilation of
cross section measurements for the elastic $\JPSI$ cross section is shown
as a function of $\Wgp$. 
The LE data from the present analysis close the gap to data from
fixed target experiments\footnote{The data from~\cite{Binkley:1981kv} and~\cite{Denby:1983az} 
have been updated using recent measurements of branching ratios~\cite{Beringer:1900zz}. The
data from~\cite{Binkley:1981kv} are also corrected 
for contributions from inelastic processes, see~\cite{webmaterial} for more 
details.}~\cite{Binkley:1981kv, Denby:1983az}  at low $\Wgp$. The fixed target data exhibit
a lower normalisation and a steeper slope than observed at HERA. 
Also shown are recent results from the LHCb experiment~\cite{Aaij:2013jxj}. 
The  extrapolated fit function for the elastic $\JPsi$ cross section is 
able to describe the LCHb data points at high $\Wgp$ well.

Following~\cite{Martin:2007sb} the obtained value of $\delta$ can 
for large photon-proton centre-of-mass energies, $\Wgp \gg m_{\JPsi}$,
be related to a leading-order gluon-density parametrised as
$x \cdot g(x, \mu^2) = N \cdot x^{- \lambda}$ via $\delta_{el} \approx 4 \cdot \lambda$.
The scale of $\JPsi$ photoproduction is often taken to be $\mu^2 = 2.4 \gevsq$.
The observed value $\lambda_{\JPsi} = 0.168 \pm 0.008$ is in remarkable agreement with
$\lambda_{incl}(Q^2 = 2.5 \gevsq) = 0.166 \pm 0.006$ obtained from fits to
inclusive DIS cross sections~\cite{Aaron:2009bp}.
Skewing effects~\cite{Martin:2007sb, Favart:2005sc} are ignored in this comparison.

\begin{figure}[hhh]
\centering
\includegraphics[width=15cm] {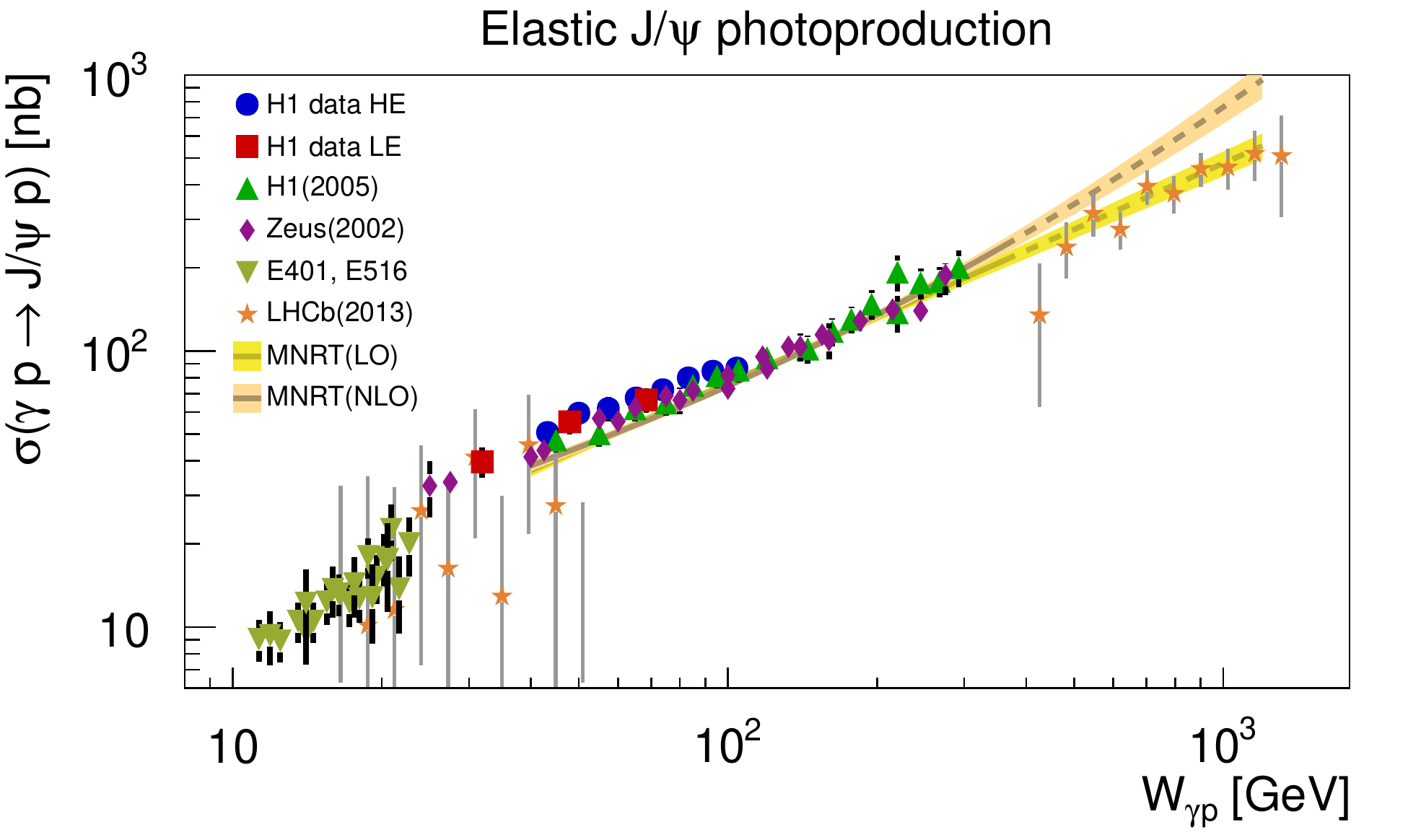}

\vspace{1cm}
\caption{
Compilation of 
elastic $\JPsi$ production cross section measurements including
this measurement, previous 
HERA results~\cite{Aktas:2005xu, Chekanov:2002xi}, 
results from fixed target experiments~\cite{Binkley:1981kv, Denby:1983az}
and from LHCb~\cite{Aaij:2013jxj}. Also presented  are QCD fits from~\cite{Martin:2007sb} to the previous 
HERA data~\cite{Aktas:2005xu, Breitweg:1997rg, Breitweg:1998nh, Chekanov:2004mw}  to determine
a gluon density at leading-order and next-to-leading order, indicated
by the curves. The fits are extrapolated in $\Wgp$ from the range of the input data to
higher values, as shown by  the dashed curves.
The shaded bands indicate the fit uncertainties.
}
\label{fig: Xsetion_WSummary_MartinFit} 
\end{figure}

In~\cite{Martin:2007sb} both a leading order and a  next-to-leading order gluon-density are 
derived, via fits to previous
$\JPsi$ measurements at HERA~\cite{Aktas:2005xu, Breitweg:1997rg, Breitweg:1998nh, Chekanov:2004mw}. 
The fit  results obtained in~\cite{Martin:2007sb} are 
compared with the data in figure~\ref{fig: Xsetion_WSummary_MartinFit}. 
Both fits are also extrapolated from the $\Wgp$ range of the input data
to higher  $\Wgp$ and compared with the LHCb measurement. The leading-order fit
describes the LHCb data well, whereas the next-to-leading order fit lies above the
LHCb cross sections.

\begin{table}[htbp]
 \centering
 \begin{tabular}{llcll}
   \toprule
   Process & Parameter & Fit value & Correlation\\
   \midrule
   \elas & $\delta_{el}$ & $0.67\pm0.03$&
   \begin{scriptsize}
  \begin{tabular}{ l @{=}r } 
  $\rho(\delta_{el}, N_{el})$& -0.08 \\
  $\rho(\delta_{el}, \delta_{pd})$& 0.01 \\
  $\rho(\delta_{el}, N_{pd})$& 0.09  
        \vspace*{2mm}\\
   \end{tabular} 
  \end{scriptsize}  \\
   				& $N_{el}$		 & $81\pm 3 \nbarn$ & 
  \begin{scriptsize}
  \begin{tabular}{ l @{=}r } 
  $\rho(N_{el}, \delta_{pd})$& -0.27 \\
  $\rho(N_{el}, N_{pd})$     & -0.18   
        \vspace*{2mm}\\
   \end{tabular} 
  \end{scriptsize}  \\
             		\pdis & $\delta_{pd}$ & $0.42\pm0.05$& 
   \begin{scriptsize}
  \begin{tabular}{ l @{=}r } 
  $\rho(\delta_{pd}, N_{pd})$& 0.09   
   \end{tabular} 
  \end{scriptsize} \\            		   
   				& $N_{pd}$		 & $66 \pm 7 \nbarn$ & \\

   \bottomrule
   
               		Ratio & $\delta_{R}  = \delta_{pd} - \delta_{el} $ & $-0.25\pm0.06$& 
   \begin{scriptsize}
  \begin{tabular}{ l @{=}r } 
  $\rho(\delta_{r}, N_{R})$& 0.14   
   \end{tabular} 
  \end{scriptsize} \\            		   
   				& $N_{R} = N_{pd}/N_{el}$		 & $0.81 \pm 0.11$ & \\

   \bottomrule
 \end{tabular}
 \caption{Parameter values obtained from the fit to the  cross sections as
 	a function of $\Wgp$,
	including their errors and correlations. 
 	The fit functions  are described in the text.
 	The parameters for the ratio of the two functions are also given. }
 \label{tab:CSFitResults_W}
\end{table}

\section{Summary}
\label{sec: Summary}
Photoproduction cross sections for elastic and proton-dissociative diffractive $\JPSI$ meson
production have been measured  as a function of $t$, the four-momentum
transfer at the proton vertex, and as a function of $\Wgp$, the 
photon proton centre-of-mass energy in the kinematic ranges 
$|t|<8 \gevsq$, $25\gev < \Wgp < 110 \gev$ and for the
proton-dissociative case $M_Y < 10 \gev$.
The data were collected in positron-proton collisions with the H1 detector at HERA,
at a centre-of-mass energy of $\sqrt{s}\approx 318 \, \gev$ and $\sqrt{s}\approx 225 \, \gev$.
Measurements in the electron and muon decay channels are combined,
and are parametrised using phenomenological fits.

The elastic and the
proton-dissociative cross sections are extracted simultaneously. Using
this technique, a precise measurement of proton-dissociative $\JPSI$ production 
was performed in the range of small $|t|$ for the first time. 
The data taken at low centre-of-mass energies close the gap
between previous H1 measurements and fixed target data. 

The data agree well with previous HERA measurements and with
a model based on two gluon exchange. The $\Wgp$-dependence of the
proton-dissociative channel is found to be significantly weaker than that 
of the elastic channel.

\section{Acknowledgements}

We are grateful to the HERA machine group whose outstanding
efforts have made this experiment possible. 
We thank the engineers and technicians for their work in constructing and
maintaining the H1 detector, our funding agencies for 
financial support, the
DESY technical staff for continual assistance
and the DESY directorate for support and for the
hospitality which they extend to the non-DESY 
members of the collaboration.
We would like to give credit to all partners contributing to the WLCG
computing infrastructure for their support for the H1 Collaboration.

\clearpage

\clearpage

\begin{table}[h]
  \centering  
  \resizebox{\linewidth}{!}{%
\input{./Tables/Table_HER_LER_W__Niced}
}  
  \caption{%
  Elastic and proton-dissociative photoproduction cross sections $\sigma\left(\mean{\Wgp^\text{bc}}\right)$ derived from the high- and low-energy
  data sets as a function of
  the photon proton centre-of-mass energy $\Wgp$
  for the processes $ep \rightarrow e \: \JPSI \: Y$, where $Y$ denotes either a proton $p$ or
  a proton-dissociative system of mass $m_p < M_Y < 10\gev$.  
  These cross sections are obtained after the combination of the
  cross sections from the  $\mu^+ \mu^-$ and $e^+ e^-$ decay channels 
  and for the phase space as defined in
  table~\ref{tab:PhaseSpace}. 
  $\mean{\Wgp^\text{bc}}$ indicates the bin centres~\cite{Lafferty:1994cj} and 
    $\Phi_\gamma^T$ is the  transverse polarised photon flux per bin.
     $\Delta_\text{tot}$ and $\Delta_\text{comb}$ denote the total and the combined statistical and channel-specific
  individual uncertainties, as obtained from the data combination, respectively.
   The global correlation coefficients $\rho_{GC}$ are also shown. The
    full covariance matrix can be found in~\cite{webmaterial}.
  The remaining columns list the bin-to-bin correlated systematic uncertainties
  corresponding to a $+1 \sigma$ shift  due to the
correlated tracking uncertainty $\delta_\text{sys}^\text{Trk,corr}$, 
the correlated triggering uncertainty $\delta_\text{sys}^\text{Trg,corr}$, 
the  uncertainty from $\psi(2S)$ contributions $\delta_\text{sys}^\text{2S}$,  
the integrated luminosities of the high- and low-energy data sets
$\delta_\text{sys}^{\mathcal{L}_H}$, $\delta_\text{sys}^{\mathcal{L}_L}$,
the tagging uncertainties in the LAr $\delta_\text{sys}^\text{LAr10}$, the plug $\delta_\text{sys}^\text{PLUG}$ and the FTS $\delta_\text{sys}^\text{FTS}$, due to the modelling of the MC $\delta_\text{sys}^\text{MC Model}$, the $Q^2$ dependance $\delta_\text{sys}^\text{\QSq}$ and the cut on the empty calorimeter $\delta_\text{sys}^\text{EC}$.
  }
  \label{tab: CS_W}
\end{table}
\begin{table}[h]
  \centering  
 \resizebox{\linewidth}{!}{%
\input{./Tables/Table_HER_t__Niced} \newline
}  
  \caption{%
  Elastic and proton-dissociative photoproduction cross sections derived from the high-energy
  data sets as a function of
  the squared four-momentum transfer at the proton vertex $t$,
  for the processes $ep \rightarrow e \JPSI \: Y$, where $Y$ denotes either a proton $p$ or
  a proton-dissociative system of mass $m_p < M_Y < 10\gev$.  
  These cross sections are obtained after the combination of the
  cross sections from the  $\mu^+ \mu^-$ and $e^+ e^-$ decay channels 
  and for the phase space as defined in
  table~\ref{tab:PhaseSpace}. 
$\mean{\T^\text{bc}}$ indicates the bin centres~\cite{Lafferty:1994cj}. 
  The  transverse polarised  photon flux $\Phi_\gamma^T$ for the given phase space range is
 $0.0953$.
 See caption of  table~\ref{tab: CS_W} for more dietails.
}
  \label{tab: CS_HER_t}
\end{table}

\begin{table}[h]
  \centering  
  \resizebox{\linewidth}{!}{%
\input{./Tables/Table_LER_t__Niced} \newline
}  
  \caption{%
Elastic and proton-dissociative photoproduction cross sections of the low-energy
  data sets as  a function of
  the squared four-momentum transfer at the proton vertex $t$,
  for the processes $ep \rightarrow \JPSI \: Y$, where $Y$ denotes either a proton $p$ or
  a proton-dissociative system of mass $M_Y > m_p$.  These cross sections are obtained after the combination of the
  cross sections from the  $\mu^+ \mu^-$ and $e^+ e^-$ decay channels 
  and for the phase space as defined in
  table~\ref{tab:PhaseSpace}. 
$\mean{\T^\text{bc}}$ indicates the bin centres~\cite{Lafferty:1994cj}. 
   The  transverse polarised  photon flux $\Phi_\gamma^T$ for the given phase space range is
 $0.1108$.
 See caption of  table~\ref{tab: CS_W} for more dietails.
  }
  \label{tab: CS_LER_t}
\end{table}



\end{document}

%% file: commands.tex


\newcommand{\T}{\ensuremath{\abs{t}}\xspace}
\newcommand{\QSq}{\ensuremath{Q^{2}}\xspace}
\newcommand{\My}{\ensuremath{M_Y}\xspace}
\newcommand{\jpsi}{\ensuremath{J/\psi}\xspace}
\newcommand{\psiTwoS}{\ensuremath{\psi(2S)}\xspace}
\newcommand{\jpsiee}{\ensuremath{\jpsi\rightarrow ee}\xspace}
\newcommand{\jpsimumu}{\ensuremath{\jpsi\rightarrow\mu\mu}\xspace}
\newcommand{\degree}{\ensuremath{{}^\circ}\xspace}
\newcommand{\abs}[1]{\ensuremath{\lvert#1\rvert}\xspace}
\newcommand{\pomeron}{\ensuremath{\mathbb{P}}\xspace}
\newcommand{\reggeon}{\ensuremath{\mathbb{R}}\xspace}
\newcommand{\mmumu}{\ensuremath{m_{\mu\mu}}\xspace}
\newcommand{\mee}{\ensuremath{m_{ee}}\xspace}

\newcommand{\norm}[1]{\ensuremath{\lVert#1\rVert}\xspace}
\newcommand{\lumi}{\ensuremath{\mathcal{L}}\xspace}
\newcommand{\BR}{\ensuremath{\mathcal{B}}\xspace}
\newcommand{\PFlux}{\ensuremath{\Phi_\gamma}\xspace}

\newcommand{\PPP}{\ensuremath{\mathbb{PPP}}\xspace}
\newcommand{\PPR}{\ensuremath{\mathbb{PPR}}\xspace}

\newcommand{\Dele}{\ensuremath{D_\text{ele}}\xspace}

\newcommand{\mean}[1]{\ensuremath{\left\langle #1\right\rangle}\xspace}
\newcommand{\pflux}{  \ensuremath{\Phi_\gamma^T}\xspace}
\newcommand{\FigLabel}[1]{%
  \sf#1
}

\renewcommand{\land}{\ensuremath{\&\!\&}\xspace}
\renewcommand{\lor}{\ensuremath{\lvert\rvert}\xspace}

\newcommand{\tagW}{\ensuremath{\text{tag}_\text{W}}\xspace}
\newcommand{\nottagW}{\ensuremath{\overline{\tagW}}\xspace}

\newcommand{\ePelas}{\ensuremath{ep\rightarrow\jpsi p}\xspace}
\newcommand{\ePpdis}{\ensuremath{ep\rightarrow\jpsi Y}\xspace}


\newcommand{\Ampl}{\ensuremath{\mathcal{A}}\xspace}

\renewcommand{\Re}{\mathfrak{Re}}
\renewcommand{\Im}{\mathfrak{Im}}

\newcommand{\valerr}[3]{%
    \ensuremath{#1^{\,+#2}_{\,-#3}}\xspace
}

\newcommand{\Eq}[1]{%
   Eq.~(\ref{#1})\xspace
}

\newcommand{\elas}{%
    \ensuremath{\gamma p\rightarrow\jpsi\,p}\xspace
}
\newcommand{\pdis}{%
    \ensuremath{\gamma p\rightarrow\jpsi\,Y}\xspace
}


%

\newcommand{\sfmath}{%
        \SetSymbolFont{operators}{normal}{\math@encoding}{\math@sfdefault}{m}{n}
        \SetSymbolFont{operators}{bold}{\math@encoding}{\math@sfdefault}{bx}{n}

}






\newcommand{\sprod}[2]{%
    \ensuremath{\left( #1 \cdot #2 \right)}\xspace
}

\renewcommand{\vec}[1]{%
  {\boldsymbol #1}
}

\newcommand{\Tfbox}[1]{%
    \raisebox{-1.2pt}{\tikz\node[draw,inner sep=1pt]{#1};}\xspace
}



%% file: h1auts.tex

C.~Alexa$^{5}$,                
V.~Andreev$^{25}$,             
A.~Baghdasaryan$^{37}$,        
S.~Baghdasaryan$^{37}$,        
W.~Bartel$^{11}$,              
K.~Begzsuren$^{34}$,           
A.~Belousov$^{25}$,            
P.~Belov$^{11}$,               
V.~Boudry$^{28}$,              
I.~Bozovic-Jelisavcic$^{2}$
% ...


\bigskip{\it
% ...
$ ^{2}$ Vinca Institute of Nuclear Sciences, University of Belgrade,
          1100 Belgrade, Serbia \\
% ...
$ ^{5}$ National Institute for Physics and Nuclear Engineering (NIPNE) ,
          Bucharest, Romania$^{ k}$ \\
% ...
$ ^{11}$ DESY, Hamburg, Germany \\
% ...
$ ^{25}$ Lebedev Physical Institute, Moscow, Russia \\
% ...
$ ^{28}$ LLR, Ecole Polytechnique, CNRS/IN2P3, Palaiseau, France \\
% ...
$ ^{34}$ Institute of Physics and Technology of the Mongolian
          Academy of Sciences, Ulaanbaatar, Mongolia \\
% ...
$ ^{37}$ Yerevan Physics Institute, Yerevan, Armenia \\
% ...
$ ^k$ Supported by the Romanian National Authority for Scientific Research
      under the contract PN 09370101 \\
% ...
}

%% file: Tables/Table_HER_LER_W__Niced.tex
\begin{tabular}{cccc c|cc cccc cccc cccc}\\
\toprule
\Wgp range & 
$\mean{\Wgp^\text{bc}}$ &
$\Phi_\gamma^T$ &
$\sigma\left(\mean{\Wgp^\text{bc}}\right)$ & 
$\Delta_\text{tot}$ &
$\Delta_\text{comb}$ & 
$\rho^\text{GC}_\text{comb}$ &
$\delta_\text{sys}^{\text{Trk,corr}}$ & 
$\delta_\text{sys}^{\text{Trg,corr}}$ & 
$\delta_\text{sys}^{\text{2S}}$ & 
$\delta_\text{sys}^{\mathcal{L}_H}$ & 
$\delta_\text{sys}^{\mathcal{L}_L}$ & 
$\delta_\text{sys}^\text{LAr10}$ &
$\delta_\text{sys}^\text{PLUG}$ & 
$\delta_\text{sys}^\text{FTS}$ & 
$\delta_\text{sys}^\text{MC Model}$ & 
$\delta_\text{sys}^\text{\QSq}$ & 
$\delta_\text{sys}^{\text{R}_\text{LT}}$ & 
$\delta_\text{sys}^\text{EC}$ \\
{[GeV]} & [GeV] & & [nb] & [nb] & [nb] &
[\%] & 
[\%] & [\%] & [\%] & [\%] & [\%] & [\%] & [\%] & [\%] & [\%] & [\%] & [\%]& [\%]\\
\midrule
\multicolumn{19}{l}{High energy data period for elastic \jpsi production}\\
40.0 - 46.5  & 43.2  & 0.0158 & 50.7 & 4.9 & 2.1 & 62 & 2.0 & 2.0 & 1.5 & 2.7 &  -  & -2.6 & 0.6 & -0.1 & -7.1 & -0.1 & 0.0 & 1.4 \\
46.5 - 53.5  & 50.0  & 0.0144 & 59.5 & 5.8 & 2.2 & 69 & 2.0 & 2.0 & 1.5 & 2.7 &  -  & -2.6 & 0.6 & -0.1 & -7.3 & -0.1 & 0.0 & 1.4 \\
53.5 - 61.2  & 57.3  & 0.0131 & 61.8 & 6.2 & 2.7 & 71 & 2.0 & 2.0 & 1.5 & 2.7 &  -  & -2.6 & 0.6 & -0.1 & -7.4 & -0.1 & 0.0 & 1.4 \\
61.2 - 69.4  & 65.3  & 0.0120 & 67.6 & 6.2 & 2.5 & 71 & 2.0 & 2.0 & 1.5 & 2.7 &  -  & -2.6 & 0.6 & -0.1 & -6.6 & -0.1 & 0.0 & 1.4 \\
69.4 - 78.4  & 73.9  & 0.0112 & 72.4 & 6.4 & 2.6 & 71 & 2.0 & 2.0 & 1.5 & 2.7 &  -  & -2.6 & 0.6 & -0.1 & -6.3 & -0.1 & 0.0 & 1.4 \\
78.4 - 88.0  & 83.2  & 0.0103 & 79.9 & 7.0 & 3.0 & 69 & 2.0 & 2.0 & 1.5 & 2.7 &  -  & -2.6 & 0.6 & -0.1 & -6.0 & -0.1 & 0.0 & 1.4 \\
88.0 - 98.5  & 93.3  & 0.0096 & 84.4 & 7.0 & 3.0 & 69 & 2.0 & 2.0 & 1.5 & 2.7 &  -  & -2.6 & 0.6 & -0.1 & -5.5 & -0.1 & 0.0 & 1.4 \\
98.5 - 110.0 & 104.3 & 0.0089 & 86.7 & 7.3 & 3.7 & 65 & 2.0 & 2.0 & 1.5 & 2.7 &  -  & -2.6 & 0.6 & -0.1 & -5.2 & -0.1 & 0.0 & 1.4 \\
\midrule
\multicolumn{19}{l}{High energy data period for proton dissociative \jpsi production}\\
40.0 - 46.5  & 43.2  & 0.0158 & 46.0 & 6.0 & 2.3 & 54 & 2.0 & 2.0 & 1.5 & 2.7 &  -  & 9.4 & -2.2 & 0.5 & -3.9 & 0.1 & 0.0 & -4.3 \\
46.5 - 53.5  & 50.0  & 0.0144 & 52.1 & 6.5 & 2.3 & 61 & 2.0 & 2.0 & 1.5 & 2.7 &  -  & 9.4 & -2.2 & 0.5 & 2.5  & 0.1 & 0.0 & -4.3 \\
53.5 - 61.2  & 57.3  & 0.0131 & 58.7 & 7.4 & 2.3 & 61 & 2.0 & 2.0 & 1.5 & 2.7 &  -  & 9.4 & -2.2 & 0.5 & 3.5  & 0.1 & 0.0 & -4.3 \\
61.2 - 69.4  & 65.3  & 0.0120 & 58.7 & 7.5 & 2.2 & 63 & 2.0 & 2.0 & 1.5 & 2.7 &  -  & 9.4 & -2.2 & 0.5 & 4.6  & 0.1 & 0.0 & -4.3 \\
69.4 - 78.4  & 73.9  & 0.0112 & 61.5 & 8.0 & 2.4 & 62 & 2.0 & 2.0 & 1.5 & 2.7 &  -  & 9.4 & -2.2 & 0.5 & 4.8  & 0.1 & 0.0 & -4.3 \\
78.4 - 88.0  & 83.2  & 0.0103 & 67.7 & 8.7 & 2.6 & 60 & 2.0 & 2.0 & 1.5 & 2.7 &  -  & 9.4 & -2.2 & 0.5 & 4.6  & 0.1 & 0.0 & -4.3 \\
88.0 - 98.5  & 93.3  & 0.0096 & 69.8 & 9.0 & 2.7 & 59 & 2.0 & 2.0 & 1.5 & 2.7 &  -  & 9.4 & -2.2 & 0.5 & 4.8  & 0.1 & 0.0 & -4.3 \\
98.5 - 110.0 & 104.2 & 0.0089 & 68.8 & 9.0 & 3.0 & 54 & 2.0 & 2.0 & 1.5 & 2.7 &  -  & 9.4 & -2.2 & 0.5 & 4.6  & 0.1 & 0.0 & -4.3 \\
\midrule
\multicolumn{19}{l}{Low energy data period for elastic \jpsi production}\\
25.0 - 39.0 & 31.9 & 0.0465 & 39.7 & 4.9 & 3.4 & 62 & 2.0 & 2.0 & 1.5 &  -  & 4.0 & -3.4 & 0.8 & -0.1 & -6.4 & -0.1 & 0.0 & 1.9 \\
39.0 - 57.0 & 47.9 & 0.0359 & 55.4 & 5.6 & 3.3 & 64 & 2.0 & 2.0 & 1.5 &  -  & 4.0 & -3.4 & 0.8 & -0.1 & -5.1 & -0.1 & 0.0 & 1.9 \\
57.0 - 80.0 & 68.4 & 0.0284 & 66.4 & 6.8 & 4.3 & 64 & 2.0 & 2.0 & 1.5 &  -  & 4.0 & -3.4 & 0.8 & -0.1 & -4.7 & -0.1 & 0.0 & 1.9 \\
\midrule
\multicolumn{19}{l}{Low energy data period for proton dissociative \jpsi production}\\
25.0 - 39.0 & 31.9 & 0.0465 & 42.0 & 8.1  & 4.8 & 59 & 2.0 & 2.0 & 1.5 &  -  & 4.0 & 12.0 & -2.9 & 0.3 & -4.6 & 0.1 & 0.0 & -6.4 \\
39.0 - 57.0 & 47.9 & 0.0359 & 55.1 & 9.4  & 4.5 & 59 & 2.0 & 2.0 & 1.5 &  -  & 4.0 & 12.0 & -2.9 & 0.3 & -2.5 & 0.1 & 0.0 & -6.4 \\
57.0 - 80.0 & 68.3 & 0.0284 & 62.0 & 10.7 & 5.3 & 57 & 2.0 & 2.0 & 1.5 &  -  & 4.0 & 12.0 & -2.9 & 0.3 & 2.5  & 0.1 & 0.0 & -6.4 \\
\bottomrule
\end{tabular}

%% file: Tables/Table_HER_t__Niced.tex
\begin{tabular}{cccc | cc cccc cccc ccc}\\
\toprule
\T range & 
$\mean{\T^\text{bc}}$ &
$\frac{d\sigma}{d\T}\left(\mean{\T^\text{bc}}\right)$ & 
$\Delta_\text{tot}$ &
$\Delta_\text{comb}$ & 
$\rho^\text{GC}_\text{comb}$ &
$\delta_\text{sys}^{\text{Trk,corr}}$ & 
$\delta_\text{sys}^{\text{Trg,corr}}$ & 
$\delta_\text{sys}^{\text{2S}}$ & 
$\delta_\text{sys}^{\mathcal{L}_H}$ & 
$\delta_\text{sys}^\text{LAr10}$ & 
$\delta_\text{sys}^\text{PLUG}$ & 
$\delta_\text{sys}^\text{FTS}$ & 
$\delta_\text{sys}^\text{MC Model}$ & 
$\delta_\text{sys}^\text{\QSq}$ & 
$\delta_\text{sys}^{\text{R}_\text{LT}}$ & 
$\delta_\text{sys}^\text{EC}$ \\
{[$\unit{GeV^2}$]} & [$\unit{GeV^2}$] & [$\unit{nb/GeV^2}$] & [$\unit{nb/GeV^2}$] & [$\unit{nb/GeV^2}$] & [\%] &
[\%] & [\%] & [\%] & [\%] & [\%] & [\%] & [\%] & [\%] & [\%] & [\%] & [\%]\\
\midrule
\multicolumn{17}{l}{High energy data period for elastic \jpsi production}\\
0.00 - 0.05 & 0.02 & 336   & 18   & 11   & 70 & 2.0 & 2.0 & 1.5 & 2.7 & -1.0  & 0.2 & -0.1 & -0.6  & -0.1 & 0.0 & 0.5  \\
0.05 - 0.11 & 0.08 & 240.5 & 12.9 & 7.2  & 71 & 2.0 & 2.0 & 1.5 & 2.7 & -1.2  & 0.3 & -0.1 & -0.7  & -0.1 & 0.0 & 0.6  \\
0.11 - 0.17 & 0.14 & 161.2 & 9.3  & 5.5  & 66 & 2.0 & 2.0 & 1.5 & 2.7 & -1.6  & 0.3 & -0.1 & -1.0  & -0.1 & 0.0 & 0.8  \\
0.17 - 0.25 & 0.21 & 111.4 & 7.0  & 4.1  & 62 & 2.0 & 2.0 & 1.5 & 2.7 & -2.2  & 0.5 & -0.1 & -1.4  & -0.1 & 0.0 & 1.0  \\
0.25 - 0.35 & 0.30 & 70.4  & 5.1  & 3.2  & 61 & 2.0 & 2.0 & 1.5 & 2.7 & -2.9  & 0.6 & -0.2 & -1.9  & -0.1 & 0.0 & 1.4  \\
0.35 - 0.49 & 0.41 & 41.2  & 3.7  & 2.2  & 59 & 2.0 & 2.0 & 1.5 & 2.7 & -4.6  & 1.0 & -0.3 & -3.0  & 0.0  & 0.0 & 2.3  \\
0.49 - 0.69 & 0.58 & 18.0  & 2.7  & 1.4  & 59 & 2.0 & 2.0 & 1.5 & 2.7 & -9.2  & 2.1 & -0.6 & -6.5  & 0.1  & 0.0 & 4.7  \\
0.69 - 1.20 & 0.90 & 4.83  & 1.75 & 0.67 & 72 & 2.0 & 2.0 & 1.5 & 2.7 & -24.0 & 5.8 & -1.4 & -18.0 & 0.8  & 0.0 & 13.0 \\
\midrule
\multicolumn{17}{l}{High energy data period for proton dissociative \jpsi production}\\
0.00 - 0.20 & 0.10 & 47.3  & 6.7   & 2.3   & 63 & 2.0 & 2.0 & 1.5 & 2.7 & 11.0 & -2.2 & 0.6 & 3.6   & -0.1 & 0.0 & -4.6 \\
0.20 - 0.40 & 0.29 & 43.8  & 6.0   & 1.9   & 64 & 2.0 & 2.0 & 1.5 & 2.7 & 11.0 & -2.4 & 0.6 & 2.2   & -0.0 & 0.0 & -4.7 \\
0.40 - 0.64 & 0.52 & 36.7  & 5.1   & 1.6   & 70 & 2.0 & 2.0 & 1.5 & 2.7 & 11.0 & -2.6 & 0.7 & 2.0   & -0.1 & 0.0 & -5.0 \\
0.64 - 0.93 & 0.78 & 27.8  & 4.2   & 1.3   & 74 & 2.0 & 2.0 & 1.5 & 2.7 & 12.0 & -2.9 & 0.7 & 2.8   & -0.1 & 0.0 & -5.7 \\
0.93 - 1.31 & 1.12 & 16.80 & 2.59  & 0.87  & 63 & 2.0 & 2.0 & 1.5 & 2.7 & 12.0 & -3.1 & 0.7 & 2.0   & -0.1 & 0.0 & -5.9 \\
1.31 - 1.83 & 1.55 & 10.05 & 1.56  & 0.52  & 49 & 2.0 & 2.0 & 1.5 & 2.7 & 12.0 & -3.1 & 0.5 & 1.7   & 0.0  & 0.0 & -6.2 \\
1.83 - 2.63 & 2.21 & 6.04  & 0.68  & 0.33  & 46 & 2.0 & 2.0 & 1.5 & 2.7 & 6.0  & -1.7 & 0.3 & -5.5  & 0.5  & 0.0 & -3.0 \\
2.63 - 4.13 & 3.30 & 2.80  & 0.38  & 0.16  & 42 & 2.0 & 2.0 & 1.5 & 2.7 & 6.7  & -1.9 & 0.3 & -8.7  & 0.5  & 0.0 & -3.6 \\
4.13 - 8.00 & 5.71 & 0.875 & 0.178 & 0.064 & 30 & 2.0 & 2.0 & 1.5 & 2.7 & 9.0  & -2.5 & 0.3 & -15.0 & 0.2  & 0.0 & -5.4 \\
\bottomrule
\end{tabular}

%% file: Tables/Table_LER_t__Niced.tex
\begin{tabular}{cccc | cc cccc cccc ccc}\\
\toprule
\T range & 
$\mean{\T^\text{bc}}$ &
$\frac{d\sigma}{d\T}\left(\mean{\T^\text{bc}}\right)$ & 
$\Delta_\text{tot}$ &
$\Delta_\text{comb}$ & 
$\rho^\text{GC}_\text{comb}$ &
$\delta_\text{sys}^{\text{Trk,corr}}$ & 
$\delta_\text{sys}^{\text{Trg,corr}}$ & 
$\delta_\text{sys}^{\text{2S}}$ & 
$\delta_\text{sys}^{\mathcal{L}_L}$ & 
$\delta_\text{sys}^\text{LAr10}$ & 
$\delta_\text{sys}^\text{PLUG}$ & 
$\delta_\text{sys}^\text{FTS}$ & 
$\delta_\text{sys}^\text{MC Model}$ & 
$\delta_\text{sys}^\text{\QSq}$ & 
$\delta_\text{sys}^{\text{R}_\text{LT}}$ & 
$\delta_\text{sys}^\text{EC}$ \\
{[$\unit{GeV^2}$]} & [$\unit{GeV^2}$] & [$\unit{nb/GeV^2}$] & [$\unit{nb/GeV^2}$] & [$\unit{nb/GeV^2}$] & [\%] &
[\%] & [\%] & [\%] & [\%] & [\%] & [\%] & [\%] & [\%] & [\%] & [\%] & [\%]\\
\midrule
\multicolumn{17}{l}{Low energy data period for elastic \jpsi production}\\
0.00 - 0.11 & 0.05 & 178  & 16  & 12  & 49 & 2.0 & 2.0 & 1.5 & 4.0 & -2.0 & 0.4 & -0.1 & -1.4 & -0.1 & 0.0 & 1.0 \\
0.11 - 0.25 & 0.17 & 99.6 & 9.4 & 7.0 & 52 & 2.0 & 2.0 & 1.5 & 4.0 & -3.0 & 0.7 & -0.1 & -1.4 & -0.1 & 0.0 & 1.5 \\
0.25 - 0.47 & 0.35 & 43.7 & 5.6 & 4.3 & 53 & 2.0 & 2.0 & 1.5 & 4.0 & -5.0 & 1.1 & -0.1 & -3.4 & -0.0 & 0.0 & 2.6 \\
0.47 - 1.20 & 0.75 & 9.7  & 1.8 & 1.3 & 57 & 2.0 & 2.0 & 1.5 & 4.0 & -9.8 & 2.2 & -0.3 & -4.8 & 0.1 & 0.0 & 5.3 \\
\midrule
\multicolumn{17}{l}{Low energy data period for proton dissociative \jpsi production}\\
0.00 - 0.50 & 0.23 & 42.8 & 7.5  & 3.5  & 63 & 2.0 & 2.0 & 1.5 & 4.0 & 13.0 & -2.9 & 0.4 & 2.1 & -0.1 & 0.0 & -6.1 \\
0.50 - 1.15 & 0.80 & 18.9 & 4.0  & 1.8  & 58 & 2.0 & 2.0 & 1.5 & 4.0 & 16.0 & -3.7 & 0.5 & -0.6 & -0.0 & 0.0 & -8.3 \\
1.15 - 2.30 & 1.67 & 8.58 & 1.54 & 0.84 & 36 & 2.0 & 2.0 & 1.5 & 4.0 & 11.0 & -2.8 & 0.2 & -5.9 & 0.2 & 0.0 & -6.2 \\
2.30 - 5.00 & 3.42 & 2.01 & 0.58 & 0.36 & 21 & 2.0 & 2.0 & 1.5 & 4.0 & 8.9 & -2.4 & 0.1 & -19.0 & 0.8 & 0.0 & -5.4 \\
\bottomrule
\end{tabular}